\documentclass[reprint,aps,prl]{revtex4-2}

\usepackage{times}
\usepackage[colorlinks=true,urlcolor=blue,citecolor=blue]{hyperref}
\usepackage{soul}
\usepackage{graphicx}
\usepackage{amsmath}

\begin{document}

	\title{Quantum Heaviside Eigen Solver}

	\author{Zheng-Zhi Sun}
	\affiliation{School of Physical Sciences, University of Chinese Academy of Sciences, P. O. Box 4588, Beijing 100049, China}

	\author{Gang Su}
	\email[Corresponding author. Email: ] {gsu@ucas.ac.cn}

	\affiliation{Kavli Institute for Theoretical Sciences, and CAS Center for Excellence in Topological Quantum Computation, University of Chinese Academy of Sciences, Beijing 100190, China}
	\affiliation{School of Physical Sciences, University of Chinese Academy of Sciences, P. O. Box 4588, Beijing 100049, China}

\begin{abstract}
	
	Solving Hamiltonian matrix is a central task in quantum many-body physics and quantum chemistry. Here we propose a novel quantum algorithm named as a quantum Heaviside eigen solver to calculate both the eigen values and eigen states of the general Hamiltonian for quantum computers. A quantum judge is suggested to determine whether all the eigen values of a given Hamiltonian is larger than a certain threshold, and the lowest eigen value with an error smaller than $\varepsilon $ can be obtained by dichotomy in $O\left( {{{\log }}{1 \over \varepsilon }} \right)$ iterations of shifting Hamiltonian and performing quantum judge. A quantum selector is proposed to calculate the corresponding eigen states. Both quantum judge and quantum selector achieve quadratic speedup from amplitude amplification over classical diagonalization methods. The present algorithm is a universal quantum eigen solver for Hamiltonian in quantum many-body systems and quantum chemistry. We test this algorithm on the quantum simulator for a physical model to show its good feasibility.
\end{abstract}

\maketitle

\section*{Introduction}
Quantum computer is a computing machine based on the principles of quantum mechanics, which preforms unitary evolutions on qubits \cite{Feynman_1982, Preskill_2018}. It promises algorithms to solve important problems with exponential or polynomial speedup for many algebraic, number theoretic and oracular algorithms \cite{Grover1996, Shor_1997, Grover_1997, Bennett_1997, Szegedy2004, Harrow_2009, Montanaro_2016}, etc. Among them, solving eigen states and eigen values of a quantum many-body system is one of central tasks in condensed matter physics and quantum chemistry \cite{Imbrie_2016, Shiozaki_2016, Kandala_2017, Jia_2018}. Many classical numerical methods were suggested to study many-body systems such as density functional theory \cite{Koch_2001}, quantum Monte Carlo \cite{Foulkes_2001}, and tensor network approach \cite{Ran_2020}, etc. However, those methods are tied up in strongly correlated systems, where representing the quantum state is classically inaccessible due to the exponential dimension of the underlying Hilbert space when the size of system is very large. This issue can be naturally avoided in a quantum computer since one can store the quantum states in a number of qubits that scale linearly with the size of the physical system. The acceleration of solving such eigen problem of Hamiltonian matrix by quantum algorithms will make great contributions to study strongly correlated electron systems \cite{LIU_2002} and to develop new materials \cite{Babbush_2018}, good catalysts \cite{Reiher_2017} and even more effective medicines \cite{Aspuru_Guzik_2018}.

At present, there are several quantum algorithms that purpose to tackle this issue. Quantum phase estimation (QPE) \cite{Cleve_1998} was proposed to estimate the eigen value of an eigen vector for a Hamiltonian matrix. This approach requires that the eigen vector \cite{Cleve_1998, Zhou_2013} or a good approximation \cite{Abrams_1999} is known. A more feasible solution is to combine a re-configurable quantum processor for the expectation estimation and a conventional or quantum computer for variational optimization \cite{Peruzzo_2014, Kandala_2017, Liu_2019, Wang_2019, Parrish_2019, Izmaylov_2019, LaRose_2019, Higgott_2019, Mitarai_2020}. This variational quantum eigen solver (VQE) designs a parameterized quantum circuit as the ground state ansatz and optimizes it according to the results from the quantum processor. A good approximation or ansatz of the ground state is necessary for QPE and VQE to find the corresponding eigen value, which confines their applications to the systems with a good understanding. Besides, adiabatic algorithms can obtain a state close to the ground state of a given Hamiltonian for sufficiently long runtimes \cite{Farhi2000}. However, the runtimes of these algorithms are extremely difficult to calculate or bound in practice, which makes adiabatic algorithms used as heuristic methods in most cases \cite{Farhi2000, Jansen_2007, Ge_2019}.

Solving the eigen problem of Hamiltonian matrix includes the calculation of eigen values and eigen states. Once either eigen values or eigen states is given, there are quantum algorithms that can calculate another one \cite{Cleve_1998, Poulin_2009, Zhou_2013, Ge_2019}. However, no digital quantum algorithm can solve both the eigen values and eigen vectors for the general form of Hamiltonian \cite{Lin2020}. An essential path to design such a quantum algorithm is to judge the difference between correct results and the trial values. More specifically, it is better to design a quantum eigen solver to judge whether all the eigen values of the given Hamiltonian are higher than a trial eigen value. Then all eigen values can be calculated with dichotomy one by one starting from the ground state energy.

\begin{figure*}[htb]
	\includegraphics[width=1\linewidth]{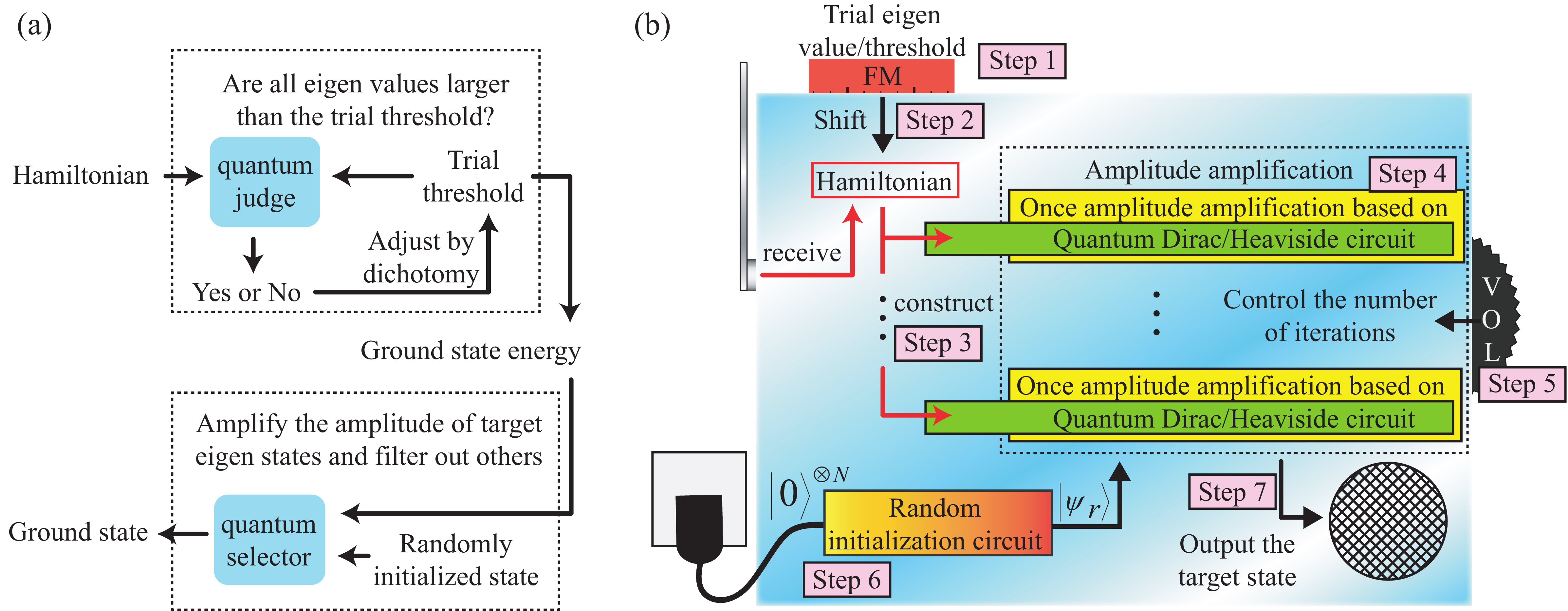}
	\caption{\label{fig-radio}Sketch of quantum Heaviside eigen solver. (a) The illustration to solve the eigen problem of the Hamiltonian with the quantum judge and quantum selector. Note that the excited state energies and the corresponding eigen states can be figured out one by one starting from the ground state energy. (b) The detailed flow-chart to design the quantum selector/judge and solve the given Hamiltonian. This process is vividly shown as a quantum radio that receives the signal (Hamiltonian) and output the contents (eigen states) according to the signal and trial frequency (eigen value/threshold). The amplitude amplification algorithm within the dash box can amplify the amplitude of states corresponding to the trial eigen value, or all eigen values lower than the given threshold, and filter out other states, which is the core of quantum Heaviside eigen solver.}
\end{figure*}

In this work we present a quantum algorithm called quantum Heaviside eigen solver (QHES) to calculate both the eigen values and eigen states of the given Hamiltonian matrix. It consists of a quantum judge to calculate the eigen values by dichotomy and a quantum selector to calculate the corresponding eigen states. The Hamiltonian is defined on the space of $N$ qubits which can be written as ${\bf{H}} = \sum\limits_l^N {{{\bf{H}}_l}} $, where each ${{{\bf{H}}_l}}$ can act on qubits \cite{Cubitt_2016}. The QHES can solve any Hamiltonian in quantum many-body systems. Generally, solving the ground state energy of this common form of Hamiltonian has been proved to be QMA-complete \cite{Kempe_2006}, where QMA stands for quantum Merlin Arthur and is the quantum analog of NP. It is generally believed that it cannot be solved in polynomial time even on quantum computer. Solving a $\chi $-dimensional Hamiltonian matrix usually requires the classical bits scaling as $O\left( \chi  \right)$ and its running time scales as $\Omega \left( \chi  \right)$ \cite{Golub_2000}. By contrast, the amplitude amplification algorithm \cite{Brassard1997, Grover_1998} with an exquisite design of oracle circuit in this work can solve this problem using the qubits that scale as $O\left( {\log \chi } \right)$ and its running time scales as $O\left( {\sqrt \chi  } \right)$.

The quantum judge is an amplitude amplification algorithm that evolves the initial state into a state belonging to the ``good subspace'', which is spanned by the eigen states with eigen values lower than the trial threshold. By detecting the presence of states in the good subspace from the output of quantum judge, one can judge whether all eigen values of the given Hamiltonian ${\bf{H}}$ are higher than the trial threshold. The lowest eigen value of ${\bf{H}}$ with an error lower than $\varepsilon $ can be obtained by $O\left( {\log {1 \over \varepsilon }} \right)$ times of binary searches of the trial threshold. Similarly, the quantum selector is an amplitude amplification algorithm where the good subspace is spanned by eigen states corresponding to the given eigen value. It can directly evolve the initial state to the target eigen states. The process of that uses QHES to solve the eigen values and eigen states is shown in Fig. \ref{fig-radio}(a). The core kernel in this process is the quantum circuit to achieve the identification of eigen states in quantum judge and quantum selector. In this paper, we accomplish this task for the general form of Hamiltonian with a quantum Heaviside circuit (QHC) and a quantum Dirac circuit (QDC).

\section*{Results}

\subsection*{A sketch of quantum Heaviside eigen solver}

The QHES is like a quantum radio in the way as shown in Fig. \ref{fig-radio}(b). This quantum radio receives the signal and outputs contents corresponding to the frequency, where the signal represents the Hamiltonian matrix, frequency represents the trial eigen value or all eigen values lower than the trial threshold, and the output contents represent the corresponding eigen states that are called qualified states. When the frequency is set at a trial threshold and QHC is adopted in the green boxes, this quantum radio judges whether there is any eigen state that corresponds to an eigen value lower than the trial threshold. Combined with the dichotomy, one can obtain the lowest eigen value with an error lower than $\varepsilon $ in $O\left( {\log {1 \over \varepsilon }} \right)$ iterations of performing quantum judge and adjusting trial threshold. This eigen value is then taken as the trial eigen value and the quantum radio can output the corresponding eigen states when QDC is adopted in the green boxes. These two processes are both achieved by the amplitude amplification based on QHC and QDC, respectively, which amplifies the amplitude of qualified states and filters out others from the randomly initialized state $\left| {{\psi _r}} \right\rangle $.

Now we explain the steps to construct the quantum selector or the quantum judge shown in Fig. \ref{fig-radio}(b).
\begin{description}
	\item[step 1] Select a trial eigen value for QDC, or a trial threshold for QHC by dichotomy.
	\item[step 2] Shift the Hamiltonian to fix the trial eigen value to $0$, or to fix the threshold to ${1 \over 2}$. This step is to avoid redesigning the whole circuit, where only the Hamiltonian evolution circuit needs to be changed.
	\item[step 3] Construct the QDC or QHC according to the shifted Hamiltonian, where the trial eigen value and trial threshold is fixed. This is the most difficult and important part in QHES.
	\item[step 4] Construct the amplitude amplification algorithm based on QDC or QHC, which is used to mark the qualified states.
	\item[step 5] Determine the number of iterations in amplitude amplification. This step is evaded since we use the fixed-point quantum search \cite{Yoder_2014} as the amplitude amplification algorithm.
	\item[step 6] Initialize the input state $\left| {{\psi _r}} \right\rangle $, which is a superposition of all eigen states. A random circuit is usually competent.
	\item[step 7] Amplify the amplitude of qualified states and filter out others using the circuit within the dash box.
\end{description}

Steps 1, 2 and 7 are classical routine operations which are introduced in the Supplemental Material. Step 3 is to design the quantum circuit of QHC and QDC with the given Hamiltonian, which is the core of QHES. Step 4 can be achieved by the standard flow to construct the amplitude amplification algorithm whose oracle circuit is QHC or QDC. Step 5 is to determine how many iterations in amplitude amplification algorithm are needed to amplify the amplitude of target state sufficiently. Using the fixed-point search, the number of iterations is $O\left( {\sqrt \chi  } \right)$ on the assumption that the overlap between $\left| {{\psi _r}} \right\rangle $ and target states is no less than $O\left( {{1 \over \chi }} \right)$. Step 6 is to initialize the initial state satisfying the above assumption, which can be done by a random circuit with high probability. Next, we will introduce the central idea to construct QDC and QHC. The mathematical analysis and detailed instructions for all steps are presented in the Supplemental Material.

\subsection*{Constructing quantum Heaviside circuit}

The purposes of QHC and QDC are to mark the qualified eigen states for the amplitude amplification. These two quantum circuits are unitary operators working on $N$ physical qubits, $K$ auxiliary qubits and one mark qubit. They output the results on the mark qubit while do not change the inputs on physical qubits at all time. The states on auxiliary qubits change in the process of quantum circuit while they are disregarded. We define the qualified states $\left| {{E_q}} \right\rangle $ for these two circuits as the states on $N$ physical qubits whose output on the mark qubit is $\left| 0 \right\rangle $. For QHC, the qualified states are eigen states whose corresponding eigen values are smaller than a given threshold $\theta$. 

The QHC is designed to filter out the eigen states with eigen values larger than the given threshold $\theta $ and to preserve as much proportion of the eigen states with eigen values smaller than $\theta $. This process consists of two parts. The first part is to identify the eigen values of all eigen states, and the second part is a simple filter circuit based on the eigen values. The QPE is proposed to entangle the eigen states with the binary representation of their corresponding eigen values on the auxiliary qubits \cite{Cleve_1998}. However, QPE is not capable for this task due to its uncertainty, which is also called heavy tail \cite{Poulin_2009}. Here we use three strategies on the original QPE algorithm to construct the quantum Heaviside circuit. The first is a multiple filtering scheme to ensure the quantum Heaviside circuit can definitely filter out the unqualified states. The second is a fine-tuning scheme to ensure the qualified states is definitely preserved. The last is a recycling scheme using a freezing operator to reduce the number of auxiliary qubits. To the best of our knowledge, this is the first filtering method for a general Hamiltonian.

The quantum phase estimation is proposed to calculate the eigen values of a given Hermite matrix, where the corresponding eigen state is given. The QPE works on the physical qubits initialized to the corresponding eigen state and $R$ extra qubits initialized to ${\left| 0 \right\rangle ^{ \otimes R}}$, which are called representation qubits. This algorithm does not influence the eigen state on physical qubits and changes the state on representation qubits to an approximation of the binary representation of the corresponding eigen value. The success probability of QPE to output the right (nearest) binary representation of the corresponding eigen value on representation qubits is no less than ${4 \over {{\pi ^2}}}$ \cite{Cleve_1998}. This indicates the filter on representation qubits after single QPE can filter out at least ${4 \over {{\pi ^2}}}$ of the unqualified states.

To make sure that all unqualified states are filtered well out, we adopt $Q = O\left( N \right)$ QPE circuits at the same time, which is the first strategy. Here all $Q$ QPE circuits act on the same physical qubits and different representation qubits, which means the total number of representation qubits is $Q \times R$. This multiple filtering scheme can reduce the amplitude of unqualified states to exponentially small, that is, lower than ${\left( {1 - {4 \over {{\pi ^2}}}} \right)^Q}$. Unfortunately, this method may filter out the qualified states when the gap between the eigen values and their nearest binary representations is large.

The strength of this filtering effect is determined by the accuracy of QPE circuit, which is related to the accuracy of binary representation of eigen values. Since the QPE circuit uses $R$ representation qubits, the maximum error to express the eigen values is ${\pi  \over {{2^R}}}$, and the corresponding accuracy of QPE circuit is ${{4 \over {{\pi ^2}}}}$. In the worst case, the eigen states with eigen values higher than the given threshold are retained up to ${\left( {1 - {4 \over {{\pi ^2}}}} \right)^{2Q}}$. Meanwhile, we can only guarantee that at least ${\left( {{4 \over {{\pi ^2}}}} \right)^{2Q}}$ of the eigen states with eigen values lower than the given threshold are preserved.

To solve this issue, we perform a batch of filtration for $W$ different Hamiltonian matrices in sequence, which are shifted from the given normalized Hamiltonian matrix recorded as ${{\bf{H}}_0}$. This set of Hamiltonian matrices can be expressed as
\begin{align}
	\label{eq-H-set}
	\left\{ {{{\bf{H}}_w}} \right\}:{{\bf{H}}_w} = {{\bf{H}}_0} + {w \over W}{\pi \over {{2^{R - 1}} }}.
\end{align}
When we filter all these Hamiltonian, the error to express the eigen value using $R$ representation qubits is reduced to no more than ${\pi \over {W{2^\pi }}}$ at least once. In this case, it is proved in the Supplemental Material that at least ${\left( {1 - {{{\pi ^2}} \over {2{W^2}}}} \right)^{2Q}}$ of the qualified states are preserved, which can be written as $O\left( 1 \right)$ when $W = O\left( {\sqrt N } \right)$. This is the second strategy to make sure that all qualified states are not filtered out at least once.

With these two strategies, the changed QPE circuit can efficiently filter out the unqualified states while preserve qualified states. The number of auxiliary qubits is $Q \times R$, which can also be written as $O\left( {N\log {1 \over \varepsilon }} \right)$. Here the term of $O\left( {\log {1 \over \varepsilon }} \right)$ is unavoidable since it is used for the binary representation of the eigen values, where $\varepsilon $ is the error bound of the quantum judge. However, its multiplication with $N$ makes this method impractical on the near term quantum hardware. For example, if we want to solve the eigen problem of a Hamiltonian matrix with the size of ${2^{50}} \times {2^{50}}$, it needs $50$ physical qubits to represent the physical system and about $50 \times {\log _2}\left( {{{10}^6}} \right)$ qubits to obtain the results with an error lower than ${10^{ - 6}}$. Next we introduce the freezing operator as the last strategy that can reduce the number of auxiliary qubits from $O\left( {N\log {1 \over \varepsilon }} \right)$ to 

\begin{align}
	\label{eq-qhc-num}
	O\left( {\log N + \log {1 \over \varepsilon }} \right).
\end{align}

The third strategy can be described as performing QPE $Q$ times on $R$ representation qubits instead of performing QPE once on $Q \times R$ representation qubits. This strategy is similar to the iterative quantum phase estimation (iQPE) \cite{Dob_ek_2007} that involves measurement and reuse of qubits, which means that it cannot be used as the subroutine of the QHES. So we design the unitary freezing operator to ``reset'' the auxiliary qubits after each QPE circuit instead of resetting the auxiliary qubits with measurement.

\subsection*{Constructing quantum Dirac circuit with quantum coin toss}

The task of QDC is to qualify the eigen states corresponding to a given eigen value, which can be done by a quantum coin toss. We define the flipping operator to achieve
\begin{align}
	\label{eq-uf-result}
	{{\bf{U}}_c}\left| {{E_j}} \right\rangle \left| 0 \right\rangle  = \cos \left( {{E_j}} \right)\left| {{E_j}} \right\rangle \left| 0 \right\rangle  + i\sin \left( {{E_j}} \right)\left| {{E_j}} \right\rangle \left| 1 \right\rangle ,
\end{align}
whose design is given in the Supplemental Material. Here we call the qubit to express states $\left| 0 \right\rangle $ and $\left| 1 \right\rangle $ as quantum coin. The ${{\bf{U}}_c}$ is designed to flip the state on the quantum coin according to the state on physical qubits. If we use $M$ quantum coins at the same time, the amplitude of all quantum coins remain ${\left| 0 \right\rangle ^{ \otimes M}}$ is ${\cos ^M}\left( {{E_j}} \right)$. When $M$ is large enough, ${\cos ^M}\left( {{E_j}} \right)$ can be seen as an analog of the Dirac function of ${{E_j}}$. 

To meet the requirements of being the oracle circuit of the quantum selector, the quantum coin toss should distinguish the eigen states with the given eigen value ${E_g}$ and the states with the closest eigen value. Without losing generality, here we suppose the given eigen value ${E_g}$ is zero and the gap between ${E_g}$ and its closest eigen value is $\Delta $. Detailed analysis in Supplemental Material shows that to obtain the $\varepsilon$-close eigen state corresponding to the given eigen value ${E_g}$, one needs the minimum number of quantum coins as
\begin{align}
	\label{eq-M}
	M = O\left( {{1 \over {{\Delta ^2}}}\left( {N + \log {1 \over \varepsilon }} \right)} \right) ,
\end{align}
and the error bound of ${E_g}$ is ${\varepsilon _0} = O\left( \Delta  \right)$. Under this condition, the QDC outputs ${\left| 0 \right\rangle ^{ \otimes M}}$ with an amplitude of $O\left( 1 \right)$ when the states on physical qubits correspond to an eigen value $\varepsilon _0$-close to ${E_g}$, and outputs ${\left| 0 \right\rangle ^{ \otimes M}}$ with an exponentially small amplitude in other cases. Using this QDC as the oracle circuit of amplitude amplification algorithm, the quantum selector can obtain the eigen states of the eigen values solved by quantum judge. Similar to the case of QHC, the number of quantum coins is far beyond the capabilities of current quantum hardware. Fortunately, this number can be exponentially reduced to $O\left( {\log M} \right)$ by the freezing operator.

\subsection*{Freezing operator}

To reduce the number of quantum coins, we replace the single toss of $M$ quantum coins to $M$ tosses of one quantum coin. A simple idea of designing such circuit is to perform ${{{\bf{U}}_c}}$ when the quantum coin is in the state of $\left| 0 \right\rangle $, and to perform an identity operator when the quantum coin is in the state of $\left| 1 \right\rangle $. However, the operation to complete this process is not unitary, which is forbidden by the quantum computer. We extend this non-unitary operator to a unitary operator by introducing the freezing operator and extra $K$ counting qubits. The freezing operator ${{\bf{U}}_F}$ acts on the quantum coin and $K$ counting qubits. The state on the counting qubits is regarded as a binary number $x$. For example, we record $\left| {0110} \right\rangle $ as $\left| {x = 6} \right\rangle $. Then ${{\bf{U}}_F}$ is designed to achieve
\begin{subequations}\label{eq-freezing}
	\begin{align}
		{{\bf{U}}_F}\left| x \right\rangle \left| 0 \right\rangle  = \left| x \right\rangle \left| 0 \right\rangle ,
	\end{align}
	\begin{align}
		{{\bf{U}}_F}\left| x \right\rangle \left| 1 \right\rangle  = \left| {x + 1} \right\rangle \left| 1 \right\rangle   .
	\end{align}
\end{subequations}
This can be easily done by a unitary ${{\bf{U}}_{add}} = \sum\limits_x {\left| {x + 1} \right\rangle \left\langle x \right|} $ controlled by the quantum coin. The ${{\bf{U}}_{add}}$ is an elementary arithmetic operation which can be efficiently performed \cite{Vedral_1996}. If we initialize the state on counting qubits to ${\left| 1 \right\rangle ^{ \otimes K}}$ and apply ${{\bf{U}}_F}$, we note that the first counting qubit, which is the highest order in the binary representation, will not be $\left| 1 \right\rangle $ for ${2^{K - 1}}$ times of performing ${{\bf{U}}_F}$ once the quantum coin is changed to $\left| 1 \right\rangle $.

This freezing operator can exponentially reduce the number of auxiliary qubits by controlling ${{\bf{U}}_c}$ with the first counting qubit. More specifically, the quantum coin toss can be performed on one quantum coin and $K = O\left( {\log M} \right)$ counting qubits instead of $M$ quantum coins. In $M$ times of performing ${{\bf{U}}_c}$ and ${{\bf{U}}_F}$, once the quantum coin is changed to $\left| 1 \right\rangle $, the state on the first counting qubit will not be $\left| 1 \right\rangle $ based on Eq. (\ref{eq-freezing}). No ${{\bf{U}}_c}$ will be applied to the quantum coin since it is controlled by the first counting qubit, which means the quantum coin is frozen in $\left| 1 \right\rangle $ once it is changed to $\left| 1 \right\rangle $. If the quantum coin is $\left| 0 \right\rangle $ after $M$ quantum coin tosses, it must remain $\left| 0 \right\rangle $ in each quantum coin toss, which means the amplitude of the quantum coin being $\left| 0 \right\rangle $ in the end is also ${\cos ^M}\left( {{E_j}} \right)$.

\begin{figure}[h]
	\centering
	\includegraphics[width=0.9\linewidth]{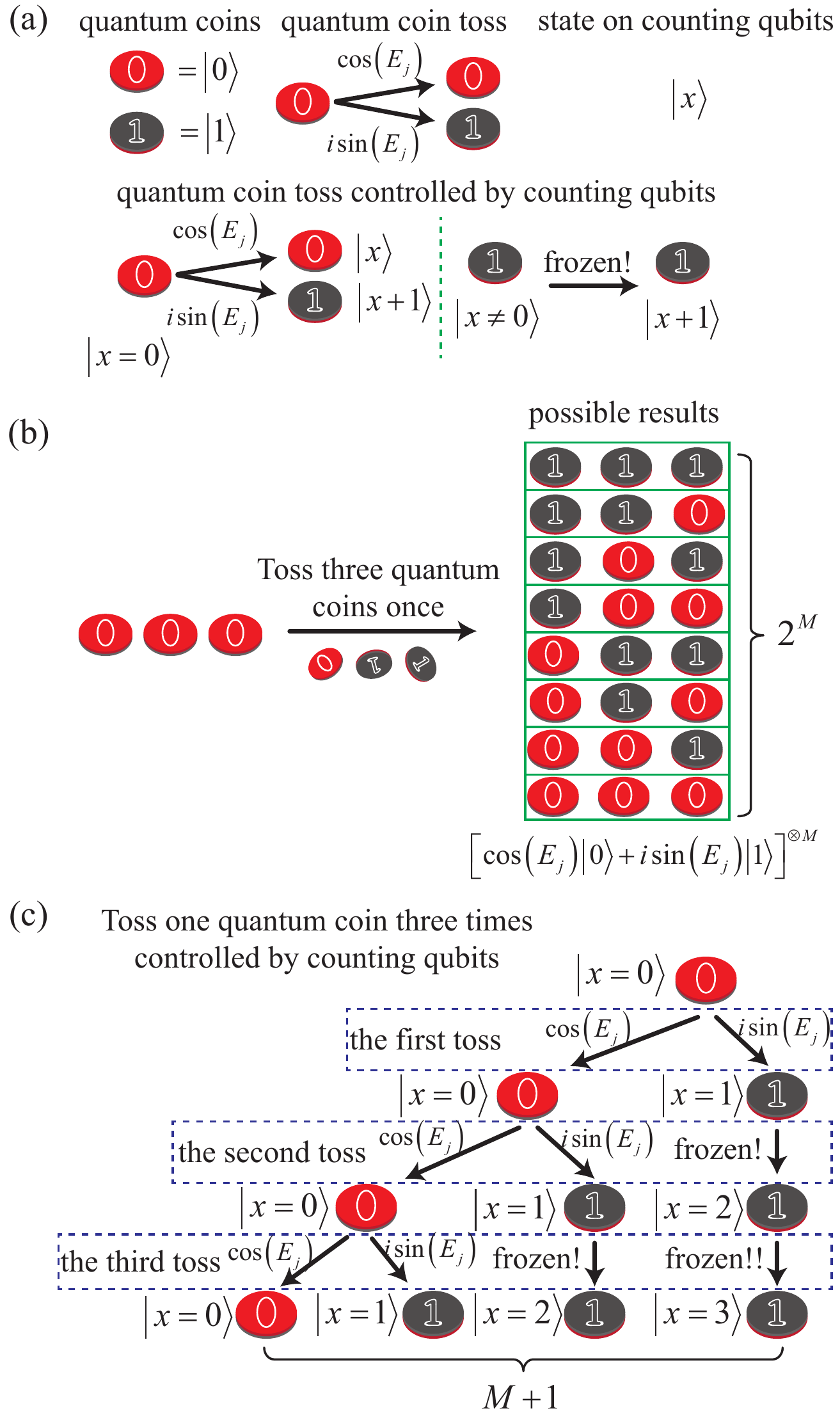}
	\caption{\label{fig-toss}Example of reducing number of quantum coins using the freezing operator. (a) Graphic representations of quantum coins, quantum coin toss, state on counting qubits and quantum coin toss controlled by counting qubits. Only coin toss processes that are used in the quantum selector are indicated. (b) All ${2^M}$ possible results of tossing $M$ quantum coins once. We need $M$ auxiliary qubits (quantum coins) to record the results. (c) The process of tossing one quantum coin $M$ times controlled by counting qubits. There are $M + 1$  possible results, which is exponentially less than the case of quantum coin toss without the control of counting qubits. We only need $O\left( {\log M} \right)$ auxiliary qubits (counting qubits) and one quantum coin to express the results. Note that there are some simplifications compared with practical design to explain this process more intuitively.}
\end{figure}

The nature of this exponential reduction of auxiliary qubits is the reduction of the dimension of the state space as shown in Fig. \ref{fig-toss}. In the quantum coin toss scheme, the quantum circuit must have the ability to express all possible results of tossing $M$ quantum coins. The number of possible results of tossing $M$ quantum coins once is ${2^M}$, which is the same as that of tossing one quantum coin $M$ times. Since we are only interested in the result of all quantum coins being $\left| 0 \right\rangle $, we choose to ignore the state with one or more quantum coins being $\left| 1 \right\rangle $. Note that we cannot decide the result of each quantum coin toss, we can only stop the quantum coin toss by the freezing operator when the result of $\left| 1 \right\rangle $ appears. The result of tossing one quantum coin $M$ times may be that all quantum coin tosses end up with $\left| 0 \right\rangle $ or one of the $M$ quantum coin tosses end up with $\left| 1 \right\rangle $. In this situation, the dimension of result space is $M + 1$, which means that we only need $O\left( {\log M} \right)$ auxiliary qubits to record the results.

In the quantum coin toss, the freezing operator guarantees that all flipping operators are performed on the $\left| 0 \right\rangle $ of the quantum coin. We define $\left| 0 \right\rangle $ as the target state of this freezing operator, where all other states is frozen. Similarly, the freezing operator can be used to reduce the number of auxiliary qubits of QHC from $O\left( {N\log {1 \over \varepsilon }} \right)$ to $O\left( {\log N + \log {1 \over \varepsilon }} \right)$. In this case, we need two freezing operators whose target states are $\left| 0 \right\rangle $ on the first (the highest order) representation qubit and ${\left| 0 \right\rangle ^{ \otimes R}}$ on all representation qubits, respectively. Note that the first representation qubit being $\left| 0 \right\rangle $ represents the eigen value being lower than ${1 \over 2}$, and ${\left| 0 \right\rangle ^{ \otimes R}}$ is the initial state of QPE algorithm. Each iteration of the filtering process starts with a QPE circuit controlled by the counting qubits, then the first freezing operator prevents the states filtered out by the QPE circuit from the following process. An inverse of the QPE circuit resets the representation qubits to an approximation of ${\left| 0 \right\rangle ^{ \otimes R}}$, and the second freezing operator makes sure that the next iteration exactly starts from ${\left| 0 \right\rangle ^{ \otimes R}}$.

With these three strategies, the QHC can effectively filter out unqualified state and preserve qualified states with only $O\left( {\log N + \log {1 \over \varepsilon }} \right)$ auxiliary qubits. Using this circuit as the oracle circuit of the amplitude amplification algorithm, the quantum judge can identify the eigen states with eigen values lower than ${1 \over 2}$ from the initial states. Since the input quantum state is randomly initialized to a superposition of all eigen states, the quantum judge can judge whether there is any eigen value of the given Hamiltonian which is lower than ${1 \over 2}$. By changing the trial threshold with dichotomy, or equivalently shifting and zooming the Hamiltonian matrix, the lowest eigen value with an error lower than $\varepsilon $ can be obtained in $O\left( {\log {1 \over \varepsilon }} \right)$ iterations. The higher eigen values can also be calculated one by one with a quantum judge that only identifies states with eigen values between two trial thresholds. After the eigen values are calculated by the quantum judge, the corresponding eigen states can be given by the quantum selector that takes quantum coin toss as the oracle circuit of the amplitude amplification algorithm. Detailed description and complexity analysis are given in the Supplemental Material.

\subsection{Simulation results of quantum judge and quantum selector}

Here we apply an open-source quantum simulator \cite{Garc_a_P_rez_2020} to produce the numerical results to solve the ground state energy of the given Hamiltonian ${\bf{H}} =  - {1 \over {N - 1}}\sum\limits_{n = 1}^{N - 1} {{\bf{\sigma }}_n^z} {\bf{\sigma }}_{n + 1}^z$ with quantum judge. The calculation of the lowest eigen value starts with a trial eigen value, and then judges whether all eigen values are larger than the trial eigen value using quantum judge. Following the standard process of dichotomy, we can obtain the lowest eigen value whose precision is confined by the precision of quantum judge. Here we define the error ${\varepsilon _v}$ to solve the ground state energy of the given Hamiltonian as
\begin{align}
	\label{eq-error-value}
	{\varepsilon _v} = \left| {{E_c} - {E_g}} \right|,
\end{align}
where ${{E_c}}$ is the result of dichotomy using quantum judge and ${{E_g} =  - 1}$ is the ground state energy of the given Hamiltonian.

\begin{figure}[htb]
	\centering
	\includegraphics[width=0.9\linewidth]{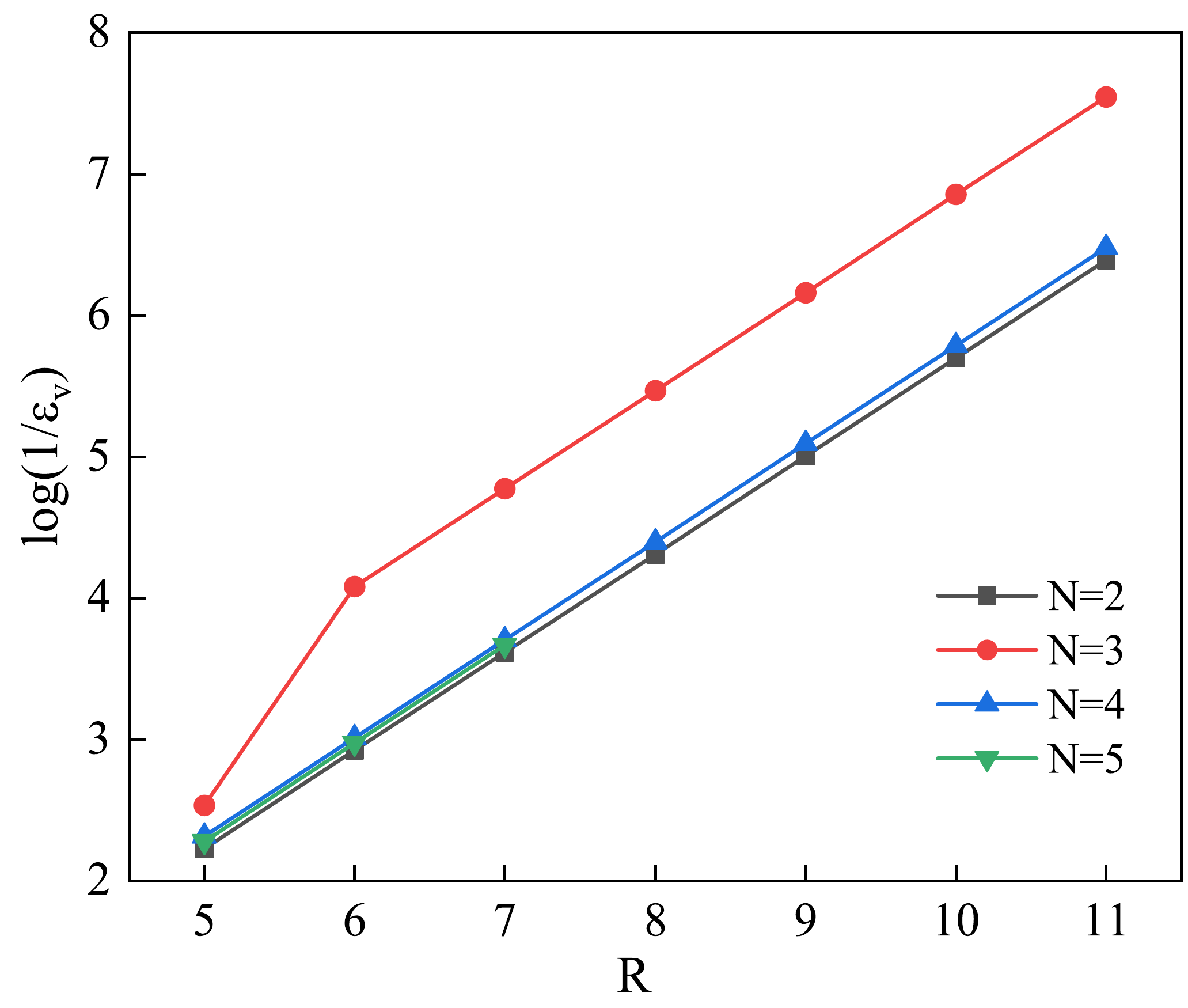}
	\caption{\label{fig-qvs}Numerical results by using quantum judge to solve the ground state energy of the given Hamiltonian ${\bf{H}} =  - {1 \over {N - 1}}\sum\limits_{n = 1}^{N - 1} {{\bf{\sigma }}_n^z} {\bf{\sigma }}_{n + 1}^z$. Here $N$ is the number of physical qubits to represent the physical system and $R$ is the number of the representation qubits in the QPE circuit. It can be seen that the error of calculated eigen value decreases exponentially with the increase of $R$. There is no result where $N = 5$ and $R > 7$ because of the limitation of our computation resource.}
\end{figure}

According to Eq. (\ref{eq-qhc-num}) and the relationship of ${\varepsilon _v} = O\left( \varepsilon  \right)$, the number of representing qubits $R$ scales linearly to $\log {1 \over \varepsilon _v}$, which is consistent with the numerical result in Fig. \ref{fig-qvs}. Due to the limitation of the depth of quantum circuits in the quantum simulator, we construct the quantum judge with an extra classical process to generate Hamiltonian set $\left\{ {{{\bf{H}}_w}} \right\}$ to make the judgment. This compromise on numerical simulation is to change the quantum search of $O\left( {\sqrt N } \right)$ Hamiltonian matrices contained in the QHC to a classical search attached to the QHC, which does not affect the verification of Eq. (\ref{eq-qhc-num}) from Fig. \ref{fig-qvs}. There is no result when $N = 5$ and $R > 7$ even with the above compromise because of the high basic cost of QPE circuit owing to the limitation of the present quantum simulator.

We also performed the numerical simulations to check the feasibility of quantum selector, which is used to solve the ground state of the given Hamiltonian ${\bf{H}} =  - {1 \over {N - 1}}\sum\limits_{n = 1}^{N - 1} {{\bf{\sigma }}_n^z} {\bf{\sigma }}_{n + 1}^z$. The ground state of this Hamiltonian is a superposition of $\left| {{\psi _0}} \right\rangle  = {\left| 0 \right\rangle ^{ \otimes N}}$ and $\left| {{\psi _1}} \right\rangle  = {\left| 1 \right\rangle ^{ \otimes N}}$. Here we define the error ${\varepsilon _s}$ to solve the ground state of the given Hamiltonian using quantum selector as
\begin{align}
	\label{eq-error-state}
	{\varepsilon _s} = 1 - {\left| {\left\langle {{{\psi _c}}}
			\mathrel{\left | {\vphantom {{{\psi _c}} {{\psi _0}}}}
				\right. \kern-\nulldelimiterspace}
			{{{\psi _0}}} \right\rangle } \right|^2} - {\left| {\left\langle {{{\psi _c}}}
			\mathrel{\left | {\vphantom {{{\psi _c}} {{\psi _1}}}}
				\right. \kern-\nulldelimiterspace}
			{{{\psi _1}}} \right\rangle } \right|^2},
\end{align}
where $\left| {{\psi _c}} \right\rangle $ is the eigen state calculated by the quantum selector. The coefficient matrix of $\left| {{\psi _c}} \right\rangle $ can be obtained by quantum tomography experimentally, or directly be outputted from the quantum simulator.

\begin{figure}[htb]
	\centering
	\includegraphics[width=0.9\linewidth]{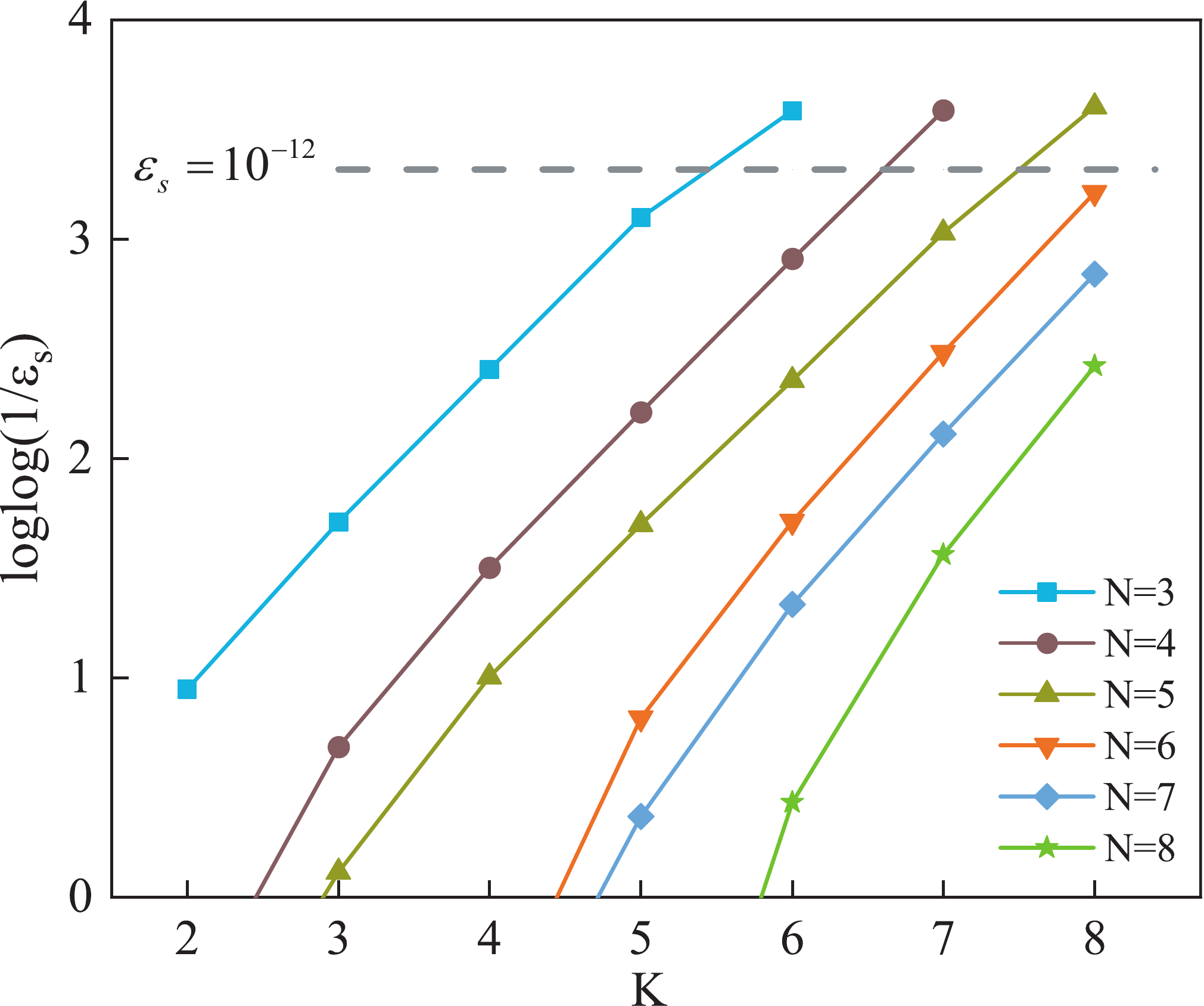}
	\caption{\label{fig-qss}Numerical results by using quantum selector to solve the ground state of the given Hamiltonian ${\bf{H}} =  - {1 \over {N - 1}}\sum\limits_{n = 1}^{N - 1} {{\bf{\sigma }}_n^z} {\bf{\sigma }}_{n + 1}^z$. Here $N$ is the number of physical qubits to represent the physical system and $K$ is the number of counting qubits. The vertical axis starts from zero where $\varepsilon $ equals approximately to $0.4$. The results with larger error are meaningless where the quantum selector does not function properly because of the lack of counting qubits. The gray dash line represents the standard line where $\varepsilon _s = {10^{ - 12}}$.}
\end{figure}

According to Eq. (\ref{eq-M}) and the relationship of ${\varepsilon _s} = O\left( {{\varepsilon ^2}} \right)$, the number of counting qubits $K$ scales linearly to $\log \log {1 \over \varepsilon _s}$, which is consistent with the numerical result in Fig. \ref{fig-qss}. It can be seen that the error of the quantum selector as quantum eigen state solver drops below ${10^{ - 12}}$ quickly. Note that we choose the simple Hamiltonian matrix to reduce the computational source required for the simulations. The eigen values and the corresponding eigen states of any k-local Hamiltonian \cite{Cubitt_2016} can be calculated using QHES. More specifically, what we only require here is the controlled time evolution of the given Hamiltonian can be efficiently implemented for unit time, which is a trivial task for any k-local Hamiltonian \cite{Berry_2015, Berry_2015_10, Low_2017, Low_2019}. The k-local Hamiltonian describes more general quantum systems than quantum many-body systems, where the former includes not only short-range interactions but also long-range interactions.

\section*{Discussion}

In this work we presented a quantum Heaviside eigen solver to solve both the eigen values and eigen states for the general Hamiltonian matrix using quantum computers. The QHES can solve the eigen values of given Hamiltonian with an error smaller than $\varepsilon $ in $O\left( {{{\log }}{1 \over \varepsilon }} \right)$ binary searches using quantum judge, which can judge whether all eigen values of the given Hamiltonian are higher than a trial threshold. Then the quantum selector in QHES outputs eigen states corresponding to the solved eigen values. In designing the oracle circuit of quantum judge, the parallel identification of eigen values is based on QPE, where three strategies are adopted to avoid its heavy tail. Besides QPE, we only use elementary quantum arithmetic operations, which is distinct from those based on non-trivial tasks such as block-encoding of the target Hamiltonian. This eigen solver was also tested on a physical model, showing its better feasibility.

We would like to mention that the QHC and the freezing operator proposed here may contribute to other quantum algorithms. To name but a few, the QHC may be used as an activation function in quantum (deep) neural networks \cite{Killoran_2019, Zhao_2021}. Furthermore, the freezing operator can exponentially reduce the dimension of freedom space (coin space) \cite{Panahiyan_2018} and the number of auxiliary qubits, which is very useful when a quantum circuit contains several parts, such as the quantum principal component analysis \cite{Lloyd_2014} and the quantum random walks \cite{Aharonov_1993, Venegas_Andraca_2012}.

\section*{Methods}

Here we give the definition of QHC and QDC. As the oracle circuit of quantum judge, QHC should identify all eigen states whose corresponding eigen values are higher than the trial threshold $\theta $. It can be expressed as 
\begin{align}
	\label{eq-Heaviside-p}
	{{\bf{U}}_H}\left| {{E_j}} \right\rangle {\left| 0 \right\rangle ^{ \otimes K}}\left| 0 \right\rangle  = {\alpha _j}\left| {{E_j}} \right\rangle {\left| 0 \right\rangle ^{ \otimes K}}\left| 0 \right\rangle +   {\beta _j}\left| {{E_j} } \right\rangle \left| {{\varphi _j}} \right\rangle \left| 1 \right\rangle ,
\end{align}
where ${\left| {{\alpha _j}} \right|^2} + {\left| {{\beta _j}} \right|^2} = 1$ and $\left| {\varphi _j} \right\rangle $ represents the state that varies with different designs of QHC on $K$ auxiliary qubits. The state $\left| {\varphi _j} \right\rangle $ on auxiliary qubits is not concerned since it does not influence the qualification of eigen states. The results of this qualification are obtained from $\left| 0 \right\rangle $ and $\left| 1 \right\rangle $ on the mark qubit. More specifically, $\left| 0 \right\rangle $ on the mark qubit is entangled with the eigen states with eigen values lower than $\theta $ and $\left| 1 \right\rangle $ is entangled with other states. To achieve this purpose, ${\alpha _j}$ should be an approximation to a Heaviside function as ${\alpha _j} = u\left(\theta - {{E_j} } \right)$, where $u$ is the classical unit step function.

Similarly, the QDC should identify eigen states whose corresponding eigen value is $\left| {{E_g}} \right\rangle $, which is defined as
\begin{align}
	\label{eq-Dirac-p}
	{{\bf{U}}_D}\left| {{E_j}} \right\rangle {\left| 0 \right\rangle ^{ \otimes K}}\left| 0 \right\rangle  = {\gamma _j}\left| {{E_j}} \right\rangle {\left| 0 \right\rangle ^{ \otimes K}} \left| 0 \right\rangle  + {\rm{ }}{\eta _j}\left| {{E_j}} \right\rangle \left| {\phi _j} \right\rangle \left| 1 \right\rangle  ,
\end{align}
where ${\left| {{\gamma _j}} \right|^2} + {\left| {{\eta _j}} \right|^2} = 1$ and $\left| {\phi _j} \right\rangle $ is also the unconcerned auxiliary state. Here ${\gamma _j}$ should be much higher when ${E_j} = {E_g}$ than when ${E_j} \ne {E_g}$, which is an analog to the Dirac function.

Detailed analysis in the Supplemental Material shows the requirements for ${\alpha _j}$ and ${\gamma _j}$ to obtain the eigen values and eigen states with an error lower than $\varepsilon $ are
\begin{equation}\label{eq-Heaviside-restrain}
	\left| {{\alpha _j}} \right|
	\begin{cases}
		 = O\left( 1 \right) & {E_j} < \theta  - \varepsilon  , \\
		\le O\left( {{1 \over \chi }} \right) & {E_j} > \theta  ,
	\end{cases}
\end{equation}
and
\begin{equation}\label{eq-restrain-alpha}
	\left| {{\gamma _j}} \right|
	\begin{cases}
		 = O\left( 1 \right) & {E_j} = {E_g}, \\
		\le O\left( {{\varepsilon  \over {\sqrt \chi  }}} \right) & {E_j} \ne {E_g},
	\end{cases}
\end{equation}
respectively.

\section*{Acknowledgments}

We thank Zi-Yong Ge for inspiring discussions. {\bf Funding:} This work is supported in part by the National Natural Science Foundation of China (11834014), the Strategic Priority Research Program of the Chinese Academy of Sciences (XDB28000000), the National Key R$\&$D Program of China (2018YFA0305800), and Beijing Municipal Science and Technology Commission (Grant No. Z190011). {\bf Author contributions:} Zheng-Zhi Sun performed all the work presented in this manuscript under the full guidance and supervision of Gang Su. {\bf Competing interests:} We declare no competing interests. {\bf Data availability:} All data needed to evaluate the conclusions in the paper are present in the paper. The code to generate the results in this paper can be obtained by reasonable request to the authors.


\begin{thebibliography}{55}%
	\makeatletter
	\providecommand \@ifxundefined [1]{%
		\@ifx{#1\undefined}
	}%
	\providecommand \@ifnum [1]{%
		\ifnum #1\expandafter \@firstoftwo
		\else \expandafter \@secondoftwo
		\fi
	}%
	\providecommand \@ifx [1]{%
		\ifx #1\expandafter \@firstoftwo
		\else \expandafter \@secondoftwo
		\fi
	}%
	\providecommand \natexlab [1]{#1}%
	\providecommand \enquote  [1]{``#1''}%
	\providecommand \bibnamefont  [1]{#1}%
	\providecommand \bibfnamefont [1]{#1}%
	\providecommand \citenamefont [1]{#1}%
	\providecommand \href@noop [0]{\@secondoftwo}%
	\providecommand \href [0]{\begingroup \@sanitize@url \@href}%
	\providecommand \@href[1]{\@@startlink{#1}\@@href}%
	\providecommand \@@href[1]{\endgroup#1\@@endlink}%
	\providecommand \@sanitize@url [0]{\catcode `\\12\catcode `\$12\catcode
		`\&12\catcode `\#12\catcode `\^12\catcode `\_12\catcode `\%12\relax}%
	\providecommand \@@startlink[1]{}%
	\providecommand \@@endlink[0]{}%
	\providecommand \url  [0]{\begingroup\@sanitize@url \@url }%
	\providecommand \@url [1]{\endgroup\@href {#1}{\urlprefix }}%
	\providecommand \urlprefix  [0]{URL }%
	\providecommand \Eprint [0]{\href }%
	\providecommand \doibase [0]{https://doi.org/}%
	\providecommand \selectlanguage [0]{\@gobble}%
	\providecommand \bibinfo  [0]{\@secondoftwo}%
	\providecommand \bibfield  [0]{\@secondoftwo}%
	\providecommand \translation [1]{[#1]}%
	\providecommand \BibitemOpen [0]{}%
	\providecommand \bibitemStop [0]{}%
	\providecommand \bibitemNoStop [0]{.\EOS\space}%
	\providecommand \EOS [0]{\spacefactor3000\relax}%
	\providecommand \BibitemShut  [1]{\csname bibitem#1\endcsname}%
	\let\auto@bib@innerbib\@empty
	\bibitem [{\citenamefont {Feynman}(1982)}]{Feynman_1982}%
	\BibitemOpen
	\bibfield  {author} {\bibinfo {author} {\bibfnamefont {R.~P.}\ \bibnamefont
			{Feynman}},\ }\bibfield  {title} {\bibinfo {title} {Simulating physics with
			computers},\ }\href {https://doi.org/https://doi.org/10.1007/BF02650179}
	{\bibfield  {journal} {\bibinfo  {journal} {International Journal of
				Theoretical Physics}\ }\textbf {\bibinfo {volume} {21}},\ \bibinfo {pages}
		{467} (\bibinfo {year} {1982})}\BibitemShut {NoStop}%
	\bibitem [{\citenamefont {Preskill}(2018)}]{Preskill_2018}%
	\BibitemOpen
	\bibfield  {author} {\bibinfo {author} {\bibfnamefont {J.}~\bibnamefont
			{Preskill}},\ }\bibfield  {title} {\bibinfo {title} {Quantum computing in the
			{NISQ} era and beyond},\ }\href {https://doi.org/10.22331/q-2018-08-06-79}
	{\bibfield  {journal} {\bibinfo  {journal} {Quantum}\ }\textbf {\bibinfo
			{volume} {2}},\ \bibinfo {pages} {79} (\bibinfo {year} {2018})}\BibitemShut
	{NoStop}%
	\bibitem [{\citenamefont {Grover}(1996)}]{Grover1996}%
	\BibitemOpen
	\bibfield  {author} {\bibinfo {author} {\bibfnamefont {L.~K.}\ \bibnamefont
			{Grover}},\ }\bibfield  {title} {\bibinfo {title} {A fast quantum mechanical
			algorithm for database search},\ }in\ \href
	{https://doi.org/10.1145/237814.237866} {\emph {\bibinfo {booktitle}
			{Proceedings of the Twenty-Eighth Annual ACM Symposium on Theory of
				Computing}}}\ (\bibinfo  {publisher} {{ACM} Press},\ \bibinfo {year}
	{1996})\BibitemShut {NoStop}%
	\bibitem [{\citenamefont {Shor}(1997)}]{Shor_1997}%
	\BibitemOpen
	\bibfield  {author} {\bibinfo {author} {\bibfnamefont {P.~W.}\ \bibnamefont
			{Shor}},\ }\bibfield  {title} {\bibinfo {title} {Polynomial-time algorithms
			for prime factorization and discrete logarithms on a quantum computer},\
	}\href {https://doi.org/10.1137/S0097539795293172} {\bibfield  {journal}
		{\bibinfo  {journal} {{SIAM} Journal on Computing}\ }\textbf {\bibinfo
			{volume} {26}},\ \bibinfo {pages} {1484} (\bibinfo {year}
		{1997})}\BibitemShut {NoStop}%
	\bibitem [{\citenamefont {Grover}(1997)}]{Grover_1997}%
	\BibitemOpen
	\bibfield  {author} {\bibinfo {author} {\bibfnamefont {L.~K.}\ \bibnamefont
			{Grover}},\ }\bibfield  {title} {\bibinfo {title} {Quantum mechanics helps in
			searching for a needle in a haystack},\ }\href
	{https://doi.org/10.1103/physrevlett.79.325} {\bibfield  {journal} {\bibinfo
			{journal} {Physical Review Letters}\ }\textbf {\bibinfo {volume} {79}},\
		\bibinfo {pages} {325} (\bibinfo {year} {1997})}\BibitemShut {NoStop}%
	\bibitem [{\citenamefont {Bennett}\ \emph {et~al.}(1997)\citenamefont
		{Bennett}, \citenamefont {Bernstein}, \citenamefont {Brassard},\ and\
		\citenamefont {Vazirani}}]{Bennett_1997}%
	\BibitemOpen
	\bibfield  {author} {\bibinfo {author} {\bibfnamefont {C.~H.}\ \bibnamefont
			{Bennett}}, \bibinfo {author} {\bibfnamefont {E.}~\bibnamefont {Bernstein}},
		\bibinfo {author} {\bibfnamefont {G.}~\bibnamefont {Brassard}},\ and\
		\bibinfo {author} {\bibfnamefont {U.}~\bibnamefont {Vazirani}},\ }\bibfield
	{title} {\bibinfo {title} {Strengths and weaknesses of quantum computing},\
	}\href {https://doi.org/10.1137/s0097539796300933} {\bibfield  {journal}
		{\bibinfo  {journal} {{SIAM} Journal on Computing}\ }\textbf {\bibinfo
			{volume} {26}},\ \bibinfo {pages} {1510} (\bibinfo {year}
		{1997})}\BibitemShut {NoStop}%
	\bibitem [{\citenamefont {Szegedy}(2004)}]{Szegedy2004}%
	\BibitemOpen
	\bibfield  {author} {\bibinfo {author} {\bibfnamefont {M.}~\bibnamefont
			{Szegedy}},\ }\bibfield  {title} {\bibinfo {title} {Quantum speed-up of
			markov chain based algorithms},\ }in\ \href
	{https://doi.org/10.1109/focs.2004.53} {\emph {\bibinfo {booktitle} {45th
				Annual {IEEE} Symposium on Foundations of Computer Science}}}\ (\bibinfo
	{publisher} {{IEEE}},\ \bibinfo {year} {2004})\BibitemShut {NoStop}%
	\bibitem [{\citenamefont {Harrow}\ \emph {et~al.}(2009)\citenamefont {Harrow},
		\citenamefont {Hassidim},\ and\ \citenamefont {Lloyd}}]{Harrow_2009}%
	\BibitemOpen
	\bibfield  {author} {\bibinfo {author} {\bibfnamefont {A.~W.}\ \bibnamefont
			{Harrow}}, \bibinfo {author} {\bibfnamefont {A.}~\bibnamefont {Hassidim}},\
		and\ \bibinfo {author} {\bibfnamefont {S.}~\bibnamefont {Lloyd}},\ }\bibfield
	{title} {\bibinfo {title} {Quantum algorithm for linear systems of
			equations},\ }\href {https://doi.org/10.1103/physrevlett.103.150502}
	{\bibfield  {journal} {\bibinfo  {journal} {Physical Review Letters}\
		}\textbf {\bibinfo {volume} {103}},\ \bibinfo {pages} {150502} (\bibinfo
		{year} {2009})}\BibitemShut {NoStop}%
	\bibitem [{\citenamefont {Montanaro}(2016)}]{Montanaro_2016}%
	\BibitemOpen
	\bibfield  {author} {\bibinfo {author} {\bibfnamefont {A.}~\bibnamefont
			{Montanaro}},\ }\bibfield  {title} {\bibinfo {title} {Quantum algorithms: an
			overview},\ }\href {https://doi.org/10.1038/npjqi.2015.23} {\bibfield
		{journal} {\bibinfo  {journal} {npj Quantum Information}\ }\textbf {\bibinfo
			{volume} {2}},\ \bibinfo {pages} {15023} (\bibinfo {year}
		{2016})}\BibitemShut {NoStop}%
	\bibitem [{\citenamefont {Imbrie}(2016)}]{Imbrie_2016}%
	\BibitemOpen
	\bibfield  {author} {\bibinfo {author} {\bibfnamefont {J.~Z.}\ \bibnamefont
			{Imbrie}},\ }\bibfield  {title} {\bibinfo {title} {On many-body localization
			for quantum spin chains},\ }\href {https://doi.org/10.1007/s10955-016-1508-x}
	{\bibfield  {journal} {\bibinfo  {journal} {Journal of Statistical Physics}\
		}\textbf {\bibinfo {volume} {163}},\ \bibinfo {pages} {998} (\bibinfo {year}
		{2016})}\BibitemShut {NoStop}%
	\bibitem [{\citenamefont {Shiozaki}(2016)}]{Shiozaki_2016}%
	\BibitemOpen
	\bibfield  {author} {\bibinfo {author} {\bibfnamefont {T.}~\bibnamefont
			{Shiozaki}},\ }\bibfield  {title} {\bibinfo {title} {An efficient solver for
			large structured eigenvalue problems in relativistic quantum chemistry},\
	}\href {https://doi.org/10.1080/00268976.2016.1158423} {\bibfield  {journal}
		{\bibinfo  {journal} {Molecular Physics}\ }\textbf {\bibinfo {volume}
			{115}},\ \bibinfo {pages} {5} (\bibinfo {year} {2016})}\BibitemShut {NoStop}%
	\bibitem [{\citenamefont {Kandala}\ \emph {et~al.}(2017)\citenamefont
		{Kandala}, \citenamefont {Mezzacapo}, \citenamefont {Temme}, \citenamefont
		{Takita}, \citenamefont {Brink}, \citenamefont {Chow},\ and\ \citenamefont
		{Gambetta}}]{Kandala_2017}%
	\BibitemOpen
	\bibfield  {author} {\bibinfo {author} {\bibfnamefont {A.}~\bibnamefont
			{Kandala}}, \bibinfo {author} {\bibfnamefont {A.}~\bibnamefont {Mezzacapo}},
		\bibinfo {author} {\bibfnamefont {K.}~\bibnamefont {Temme}}, \bibinfo
		{author} {\bibfnamefont {M.}~\bibnamefont {Takita}}, \bibinfo {author}
		{\bibfnamefont {M.}~\bibnamefont {Brink}}, \bibinfo {author} {\bibfnamefont
			{J.~M.}\ \bibnamefont {Chow}},\ and\ \bibinfo {author} {\bibfnamefont
			{J.~M.}\ \bibnamefont {Gambetta}},\ }\bibfield  {title} {\bibinfo {title}
		{Hardware-efficient variational quantum eigensolver for small molecules and
			quantum magnets},\ }\href {https://doi.org/10.1038/nature23879} {\bibfield
		{journal} {\bibinfo  {journal} {Nature}\ }\textbf {\bibinfo {volume} {549}},\
		\bibinfo {pages} {242} (\bibinfo {year} {2017})}\BibitemShut {NoStop}%
	\bibitem [{\citenamefont {Jia}\ \emph {et~al.}(2018)\citenamefont {Jia},
		\citenamefont {Wang}, \citenamefont {Mendl}, \citenamefont {Moritz},\ and\
		\citenamefont {Devereaux}}]{Jia_2018}%
	\BibitemOpen
	\bibfield  {author} {\bibinfo {author} {\bibfnamefont {C.}~\bibnamefont
			{Jia}}, \bibinfo {author} {\bibfnamefont {Y.}~\bibnamefont {Wang}}, \bibinfo
		{author} {\bibfnamefont {C.}~\bibnamefont {Mendl}}, \bibinfo {author}
		{\bibfnamefont {B.}~\bibnamefont {Moritz}},\ and\ \bibinfo {author}
		{\bibfnamefont {T.}~\bibnamefont {Devereaux}},\ }\bibfield  {title} {\bibinfo
		{title} {Paradeisos: A perfect hashing algorithm for many-body eigenvalue
			problems},\ }\href {https://doi.org/10.1016/j.cpc.2017.11.011} {\bibfield
		{journal} {\bibinfo  {journal} {Computer Physics Communications}\ }\textbf
		{\bibinfo {volume} {224}},\ \bibinfo {pages} {81} (\bibinfo {year}
		{2018})}\BibitemShut {NoStop}%
	\bibitem [{\citenamefont {Koch}\ and\ \citenamefont
		{Holthausen}(2001)}]{Koch_2001}%
	\BibitemOpen
	\bibfield  {author} {\bibinfo {author} {\bibfnamefont {W.}~\bibnamefont
			{Koch}}\ and\ \bibinfo {author} {\bibfnamefont {M.~C.}\ \bibnamefont
			{Holthausen}},\ }\href {https://doi.org/10.1002/3527600043} {\emph {\bibinfo
			{title} {A Chemist’s Guide to Density Functional Theory}}}\ (\bibinfo
	{publisher} {Wiley},\ \bibinfo {address} {New York},\ \bibinfo {year}
	{2001})\BibitemShut {NoStop}%
	\bibitem [{\citenamefont {Foulkes}\ \emph {et~al.}(2001)\citenamefont
		{Foulkes}, \citenamefont {Mitas}, \citenamefont {Needs},\ and\ \citenamefont
		{Rajagopal}}]{Foulkes_2001}%
	\BibitemOpen
	\bibfield  {author} {\bibinfo {author} {\bibfnamefont {W.~M.~C.}\
			\bibnamefont {Foulkes}}, \bibinfo {author} {\bibfnamefont {L.}~\bibnamefont
			{Mitas}}, \bibinfo {author} {\bibfnamefont {R.~J.}\ \bibnamefont {Needs}},\
		and\ \bibinfo {author} {\bibfnamefont {G.}~\bibnamefont {Rajagopal}},\
	}\bibfield  {title} {\bibinfo {title} {Quantum monte carlo simulations of
			solids},\ }\href {https://doi.org/10.1103/revmodphys.73.33} {\bibfield
		{journal} {\bibinfo  {journal} {Reviews of Modern Physics}\ }\textbf
		{\bibinfo {volume} {73}},\ \bibinfo {pages} {33} (\bibinfo {year}
		{2001})}\BibitemShut {NoStop}%
	\bibitem [{\citenamefont {Ran}\ \emph {et~al.}(2020)\citenamefont {Ran},
		\citenamefont {Tirrito}, \citenamefont {Peng}, \citenamefont {Chen},
		\citenamefont {Tagliacozzo}, \citenamefont {Su},\ and\ \citenamefont
		{Lewenstein}}]{Ran_2020}%
	\BibitemOpen
	\bibfield  {author} {\bibinfo {author} {\bibfnamefont {S.-J.}\ \bibnamefont
			{Ran}}, \bibinfo {author} {\bibfnamefont {E.}~\bibnamefont {Tirrito}},
		\bibinfo {author} {\bibfnamefont {C.}~\bibnamefont {Peng}}, \bibinfo {author}
		{\bibfnamefont {X.}~\bibnamefont {Chen}}, \bibinfo {author} {\bibfnamefont
			{L.}~\bibnamefont {Tagliacozzo}}, \bibinfo {author} {\bibfnamefont
			{G.}~\bibnamefont {Su}},\ and\ \bibinfo {author} {\bibfnamefont
			{M.}~\bibnamefont {Lewenstein}},\ }\href
	{https://doi.org/10.1007/978-3-030-34489-4} {\emph {\bibinfo {title} {Tensor
				Network Contractions}}}\ (\bibinfo  {publisher} {Springer International
		Publishing},\ \bibinfo {address} {Berlin},\ \bibinfo {year}
	{2020})\BibitemShut {NoStop}%
	\bibitem [{\citenamefont {Liu}(2002)}]{LIU_2002}%
	\BibitemOpen
	\bibfield  {author} {\bibinfo {author} {\bibfnamefont {Y.-L.}\ \bibnamefont
			{Liu}},\ }\bibfield  {title} {\bibinfo {title} {Universal description of
			strongly correlated systems},\ }\href
	{https://doi.org/10.1142/s0217979202009949} {\bibfield  {journal} {\bibinfo
			{journal} {International Journal of Modern Physics B}\ }\textbf {\bibinfo
			{volume} {16}},\ \bibinfo {pages} {773} (\bibinfo {year} {2002})}\BibitemShut
	{NoStop}%
	\bibitem [{\citenamefont {Babbush}\ \emph {et~al.}(2018)\citenamefont
		{Babbush}, \citenamefont {Gidney}, \citenamefont {Berry}, \citenamefont
		{Wiebe}, \citenamefont {McClean}, \citenamefont {Paler}, \citenamefont
		{Fowler},\ and\ \citenamefont {Neven}}]{Babbush_2018}%
	\BibitemOpen
	\bibfield  {author} {\bibinfo {author} {\bibfnamefont {R.}~\bibnamefont
			{Babbush}}, \bibinfo {author} {\bibfnamefont {C.}~\bibnamefont {Gidney}},
		\bibinfo {author} {\bibfnamefont {D.~W.}\ \bibnamefont {Berry}}, \bibinfo
		{author} {\bibfnamefont {N.}~\bibnamefont {Wiebe}}, \bibinfo {author}
		{\bibfnamefont {J.}~\bibnamefont {McClean}}, \bibinfo {author} {\bibfnamefont
			{A.}~\bibnamefont {Paler}}, \bibinfo {author} {\bibfnamefont
			{A.}~\bibnamefont {Fowler}},\ and\ \bibinfo {author} {\bibfnamefont
			{H.}~\bibnamefont {Neven}},\ }\bibfield  {title} {\bibinfo {title} {Encoding
			electronic spectra in quantum circuits with linear t complexity},\ }\href
	{https://doi.org/10.1103/physrevx.8.041015} {\bibfield  {journal} {\bibinfo
			{journal} {Physical Review X}\ }\textbf {\bibinfo {volume} {8}},\ \bibinfo
		{pages} {041015} (\bibinfo {year} {2018})}\BibitemShut {NoStop}%
	\bibitem [{\citenamefont {Reiher}\ \emph {et~al.}(2017)\citenamefont {Reiher},
		\citenamefont {Wiebe}, \citenamefont {Svore}, \citenamefont {Wecker},\ and\
		\citenamefont {Troyer}}]{Reiher_2017}%
	\BibitemOpen
	\bibfield  {author} {\bibinfo {author} {\bibfnamefont {M.}~\bibnamefont
			{Reiher}}, \bibinfo {author} {\bibfnamefont {N.}~\bibnamefont {Wiebe}},
		\bibinfo {author} {\bibfnamefont {K.~M.}\ \bibnamefont {Svore}}, \bibinfo
		{author} {\bibfnamefont {D.}~\bibnamefont {Wecker}},\ and\ \bibinfo {author}
		{\bibfnamefont {M.}~\bibnamefont {Troyer}},\ }\bibfield  {title} {\bibinfo
		{title} {Elucidating reaction mechanisms on quantum computers},\ }\href
	{https://doi.org/10.1073/pnas.1619152114} {\bibfield  {journal} {\bibinfo
			{journal} {Proceedings of the National Academy of Sciences}\ }\textbf
		{\bibinfo {volume} {114}},\ \bibinfo {pages} {7555} (\bibinfo {year}
		{2017})}\BibitemShut {NoStop}%
	\bibitem [{\citenamefont {Aspuru-Guzik}\ \emph {et~al.}(2018)\citenamefont
		{Aspuru-Guzik}, \citenamefont {Lindh},\ and\ \citenamefont
		{Reiher}}]{Aspuru_Guzik_2018}%
	\BibitemOpen
	\bibfield  {author} {\bibinfo {author} {\bibfnamefont {A.}~\bibnamefont
			{Aspuru-Guzik}}, \bibinfo {author} {\bibfnamefont {R.}~\bibnamefont
			{Lindh}},\ and\ \bibinfo {author} {\bibfnamefont {M.}~\bibnamefont
			{Reiher}},\ }\bibfield  {title} {\bibinfo {title} {The matter simulation
			(r)evolution},\ }\href {https://doi.org/10.1021/acscentsci.7b00550}
	{\bibfield  {journal} {\bibinfo  {journal} {{ACS} Central Science}\ }\textbf
		{\bibinfo {volume} {4}},\ \bibinfo {pages} {144} (\bibinfo {year}
		{2018})}\BibitemShut {NoStop}%
	\bibitem [{\citenamefont {Cleve}\ \emph {et~al.}(1998)\citenamefont {Cleve},
		\citenamefont {Ekert}, \citenamefont {Macchiavello},\ and\ \citenamefont
		{Mosca}}]{Cleve_1998}%
	\BibitemOpen
	\bibfield  {author} {\bibinfo {author} {\bibfnamefont {R.}~\bibnamefont
			{Cleve}}, \bibinfo {author} {\bibfnamefont {A.}~\bibnamefont {Ekert}},
		\bibinfo {author} {\bibfnamefont {C.}~\bibnamefont {Macchiavello}},\ and\
		\bibinfo {author} {\bibfnamefont {M.}~\bibnamefont {Mosca}},\ }\bibfield
	{title} {\bibinfo {title} {Quantum algorithms revisited},\ }\href
	{https://doi.org/10.1098/rspa.1998.0164} {\bibfield  {journal} {\bibinfo
			{journal} {Proceedings of the Royal Society of London. Series A:
				Mathematical, Physical and Engineering Sciences}\ }\textbf {\bibinfo {volume}
			{454}},\ \bibinfo {pages} {339} (\bibinfo {year} {1998})}\BibitemShut
	{NoStop}%
	\bibitem [{\citenamefont {Zhou}\ \emph {et~al.}(2013)\citenamefont {Zhou},
		\citenamefont {Kalasuwan}, \citenamefont {Ralph},\ and\ \citenamefont
		{O‘Brien}}]{Zhou_2013}%
	\BibitemOpen
	\bibfield  {author} {\bibinfo {author} {\bibfnamefont {X.-Q.}\ \bibnamefont
			{Zhou}}, \bibinfo {author} {\bibfnamefont {P.}~\bibnamefont {Kalasuwan}},
		\bibinfo {author} {\bibfnamefont {T.~C.}\ \bibnamefont {Ralph}},\ and\
		\bibinfo {author} {\bibfnamefont {J.~L.}\ \bibnamefont {O‘Brien}},\
	}\bibfield  {title} {\bibinfo {title} {Calculating unknown eigenvalues with a
			quantum algorithm},\ }\href {https://doi.org/10.1038/nphoton.2012.360}
	{\bibfield  {journal} {\bibinfo  {journal} {Nature Photonics}\ }\textbf
		{\bibinfo {volume} {7}},\ \bibinfo {pages} {223} (\bibinfo {year}
		{2013})}\BibitemShut {NoStop}%
	\bibitem [{\citenamefont {Abrams}\ and\ \citenamefont
		{Lloyd}(1999)}]{Abrams_1999}%
	\BibitemOpen
	\bibfield  {author} {\bibinfo {author} {\bibfnamefont {D.~S.}\ \bibnamefont
			{Abrams}}\ and\ \bibinfo {author} {\bibfnamefont {S.}~\bibnamefont {Lloyd}},\
	}\bibfield  {title} {\bibinfo {title} {Quantum algorithm providing
			exponential speed increase for finding eigenvalues and eigenvectors},\ }\href
	{https://doi.org/10.1103/physrevlett.83.5162} {\bibfield  {journal} {\bibinfo
			{journal} {Physical Review Letters}\ }\textbf {\bibinfo {volume} {83}},\
		\bibinfo {pages} {5162} (\bibinfo {year} {1999})}\BibitemShut {NoStop}%
	\bibitem [{\citenamefont {Peruzzo}\ \emph {et~al.}(2014)\citenamefont
		{Peruzzo}, \citenamefont {McClean}, \citenamefont {Shadbolt}, \citenamefont
		{Yung}, \citenamefont {Zhou}, \citenamefont {Love}, \citenamefont
		{Aspuru-Guzik},\ and\ \citenamefont {O'Brien}}]{Peruzzo_2014}%
	\BibitemOpen
	\bibfield  {author} {\bibinfo {author} {\bibfnamefont {A.}~\bibnamefont
			{Peruzzo}}, \bibinfo {author} {\bibfnamefont {J.}~\bibnamefont {McClean}},
		\bibinfo {author} {\bibfnamefont {P.}~\bibnamefont {Shadbolt}}, \bibinfo
		{author} {\bibfnamefont {M.-H.}\ \bibnamefont {Yung}}, \bibinfo {author}
		{\bibfnamefont {X.-Q.}\ \bibnamefont {Zhou}}, \bibinfo {author}
		{\bibfnamefont {P.~J.}\ \bibnamefont {Love}}, \bibinfo {author}
		{\bibfnamefont {A.}~\bibnamefont {Aspuru-Guzik}},\ and\ \bibinfo {author}
		{\bibfnamefont {J.~L.}\ \bibnamefont {O'Brien}},\ }\bibfield  {title}
	{\bibinfo {title} {A variational eigenvalue solver on a photonic quantum
			processor},\ }\href {https://doi.org/10.1038/ncomms5213} {\bibfield
		{journal} {\bibinfo  {journal} {Nature Communications}\ }\textbf {\bibinfo
			{volume} {5}},\ \bibinfo {pages} {4213} (\bibinfo {year} {2014})}\BibitemShut
	{NoStop}%
	\bibitem [{\citenamefont {Liu}\ \emph {et~al.}(2019)\citenamefont {Liu},
		\citenamefont {Zhang}, \citenamefont {Wan},\ and\ \citenamefont
		{Wang}}]{Liu_2019}%
	\BibitemOpen
	\bibfield  {author} {\bibinfo {author} {\bibfnamefont {J.-G.}\ \bibnamefont
			{Liu}}, \bibinfo {author} {\bibfnamefont {Y.-H.}\ \bibnamefont {Zhang}},
		\bibinfo {author} {\bibfnamefont {Y.}~\bibnamefont {Wan}},\ and\ \bibinfo
		{author} {\bibfnamefont {L.}~\bibnamefont {Wang}},\ }\bibfield  {title}
	{\bibinfo {title} {Variational quantum eigensolver with fewer qubits},\
	}\href {https://doi.org/10.1103/physrevresearch.1.023025} {\bibfield
		{journal} {\bibinfo  {journal} {Physical Review Research}\ }\textbf {\bibinfo
			{volume} {1}},\ \bibinfo {pages} {023025} (\bibinfo {year}
		{2019})}\BibitemShut {NoStop}%
	\bibitem [{\citenamefont {Wang}\ \emph {et~al.}(2019)\citenamefont {Wang},
		\citenamefont {Higgott},\ and\ \citenamefont {Brierley}}]{Wang_2019}%
	\BibitemOpen
	\bibfield  {author} {\bibinfo {author} {\bibfnamefont {D.}~\bibnamefont
			{Wang}}, \bibinfo {author} {\bibfnamefont {O.}~\bibnamefont {Higgott}},\ and\
		\bibinfo {author} {\bibfnamefont {S.}~\bibnamefont {Brierley}},\ }\bibfield
	{title} {\bibinfo {title} {Accelerated variational quantum eigensolver},\
	}\href {https://doi.org/10.1103/physrevlett.122.140504} {\bibfield  {journal}
		{\bibinfo  {journal} {Physical Review Letters}\ }\textbf {\bibinfo {volume}
			{122}},\ \bibinfo {pages} {140504} (\bibinfo {year} {2019})}\BibitemShut
	{NoStop}%
	\bibitem [{\citenamefont {Parrish}\ \emph {et~al.}(2019)\citenamefont
		{Parrish}, \citenamefont {Hohenstein}, \citenamefont {McMahon},\ and\
		\citenamefont {Mart{\'{\i}}nez}}]{Parrish_2019}%
	\BibitemOpen
	\bibfield  {author} {\bibinfo {author} {\bibfnamefont {R.~M.}\ \bibnamefont
			{Parrish}}, \bibinfo {author} {\bibfnamefont {E.~G.}\ \bibnamefont
			{Hohenstein}}, \bibinfo {author} {\bibfnamefont {P.~L.}\ \bibnamefont
			{McMahon}},\ and\ \bibinfo {author} {\bibfnamefont {T.~J.}\ \bibnamefont
			{Mart{\'{\i}}nez}},\ }\bibfield  {title} {\bibinfo {title} {Quantum
			computation of electronic transitions using a variational quantum
			eigensolver},\ }\href {https://doi.org/10.1103/physrevlett.122.230401}
	{\bibfield  {journal} {\bibinfo  {journal} {Physical Review Letters}\
		}\textbf {\bibinfo {volume} {122}},\ \bibinfo {pages} {230401} (\bibinfo
		{year} {2019})}\BibitemShut {NoStop}%
	\bibitem [{\citenamefont {Izmaylov}\ \emph {et~al.}(2019)\citenamefont
		{Izmaylov}, \citenamefont {Yen}, \citenamefont {Lang},\ and\ \citenamefont
		{Verteletskyi}}]{Izmaylov_2019}%
	\BibitemOpen
	\bibfield  {author} {\bibinfo {author} {\bibfnamefont {A.~F.}\ \bibnamefont
			{Izmaylov}}, \bibinfo {author} {\bibfnamefont {T.-C.}\ \bibnamefont {Yen}},
		\bibinfo {author} {\bibfnamefont {R.~A.}\ \bibnamefont {Lang}},\ and\
		\bibinfo {author} {\bibfnamefont {V.}~\bibnamefont {Verteletskyi}},\
	}\bibfield  {title} {\bibinfo {title} {Unitary partitioning approach to the
			measurement problem in the variational quantum eigensolver method},\ }\href
	{https://doi.org/10.1021/acs.jctc.9b00791} {\bibfield  {journal} {\bibinfo
			{journal} {Journal of Chemical Theory and Computation}\ }\textbf {\bibinfo
			{volume} {16}},\ \bibinfo {pages} {190} (\bibinfo {year} {2019})}\BibitemShut
	{NoStop}%
	\bibitem [{\citenamefont {LaRose}\ \emph {et~al.}(2019)\citenamefont {LaRose},
		\citenamefont {Tikku}, \citenamefont {O'Neel-Judy}, \citenamefont {Cincio},\
		and\ \citenamefont {Coles}}]{LaRose_2019}%
	\BibitemOpen
	\bibfield  {author} {\bibinfo {author} {\bibfnamefont {R.}~\bibnamefont
			{LaRose}}, \bibinfo {author} {\bibfnamefont {A.}~\bibnamefont {Tikku}},
		\bibinfo {author} {\bibfnamefont {{\'{E}}.}~\bibnamefont {O'Neel-Judy}},
		\bibinfo {author} {\bibfnamefont {L.}~\bibnamefont {Cincio}},\ and\ \bibinfo
		{author} {\bibfnamefont {P.~J.}\ \bibnamefont {Coles}},\ }\bibfield  {title}
	{\bibinfo {title} {Variational quantum state diagonalization},\ }\href
	{https://doi.org/10.1038/s41534-019-0167-6} {\bibfield  {journal} {\bibinfo
			{journal} {npj Quantum Information}\ }\textbf {\bibinfo {volume} {5}},\
		\bibinfo {pages} {57} (\bibinfo {year} {2019})}\BibitemShut {NoStop}%
	\bibitem [{\citenamefont {Higgott}\ \emph {et~al.}(2019)\citenamefont
		{Higgott}, \citenamefont {Wang},\ and\ \citenamefont
		{Brierley}}]{Higgott_2019}%
	\BibitemOpen
	\bibfield  {author} {\bibinfo {author} {\bibfnamefont {O.}~\bibnamefont
			{Higgott}}, \bibinfo {author} {\bibfnamefont {D.}~\bibnamefont {Wang}},\ and\
		\bibinfo {author} {\bibfnamefont {S.}~\bibnamefont {Brierley}},\ }\bibfield
	{title} {\bibinfo {title} {Variational quantum computation of excited
			states},\ }\href {https://doi.org/10.22331/q-2019-07-01-156} {\bibfield
		{journal} {\bibinfo  {journal} {Quantum}\ }\textbf {\bibinfo {volume} {3}},\
		\bibinfo {pages} {156} (\bibinfo {year} {2019})}\BibitemShut {NoStop}%
	\bibitem [{\citenamefont {Mitarai}\ \emph {et~al.}(2020)\citenamefont
		{Mitarai}, \citenamefont {Nakagawa},\ and\ \citenamefont
		{Mizukami}}]{Mitarai_2020}%
	\BibitemOpen
	\bibfield  {author} {\bibinfo {author} {\bibfnamefont {K.}~\bibnamefont
			{Mitarai}}, \bibinfo {author} {\bibfnamefont {Y.~O.}\ \bibnamefont
			{Nakagawa}},\ and\ \bibinfo {author} {\bibfnamefont {W.}~\bibnamefont
			{Mizukami}},\ }\bibfield  {title} {\bibinfo {title} {Theory of analytical
			energy derivatives for the variational quantum eigensolver},\ }\href
	{https://doi.org/10.1103/physrevresearch.2.013129} {\bibfield  {journal}
		{\bibinfo  {journal} {Physical Review Research}\ }\textbf {\bibinfo {volume}
			{2}},\ \bibinfo {pages} {013129} (\bibinfo {year} {2020})}\BibitemShut
	{NoStop}%
	\bibitem [{\citenamefont {Farhi}\ \emph {et~al.}(2000)\citenamefont {Farhi},
		\citenamefont {Goldstone}, \citenamefont {Gutmann},\ and\ \citenamefont
		{Sipser}}]{Farhi2000}%
	\BibitemOpen
	\bibfield  {author} {\bibinfo {author} {\bibfnamefont {E.}~\bibnamefont
			{Farhi}}, \bibinfo {author} {\bibfnamefont {J.}~\bibnamefont {Goldstone}},
		\bibinfo {author} {\bibfnamefont {S.}~\bibnamefont {Gutmann}},\ and\ \bibinfo
		{author} {\bibfnamefont {M.}~\bibnamefont {Sipser}},\ }\bibfield  {title}
	{\bibinfo {title} {Quantum computation by adiabatic evolution},\ }\href@noop
	{} {\  (\bibinfo {year} {2000})},\ \Eprint
	{https://arxiv.org/abs/quant-ph/0001106} {arXiv:quant-ph/0001106 [quant-ph]}
	\BibitemShut {NoStop}%
	\bibitem [{\citenamefont {Jansen}\ \emph {et~al.}(2007)\citenamefont {Jansen},
		\citenamefont {Ruskai},\ and\ \citenamefont {Seiler}}]{Jansen_2007}%
	\BibitemOpen
	\bibfield  {author} {\bibinfo {author} {\bibfnamefont {S.}~\bibnamefont
			{Jansen}}, \bibinfo {author} {\bibfnamefont {M.-B.}\ \bibnamefont {Ruskai}},\
		and\ \bibinfo {author} {\bibfnamefont {R.}~\bibnamefont {Seiler}},\
	}\bibfield  {title} {\bibinfo {title} {Bounds for the adiabatic approximation
			with applications to quantum computation},\ }\href
	{https://doi.org/10.1063/1.2798382} {\bibfield  {journal} {\bibinfo
			{journal} {Journal of Mathematical Physics}\ }\textbf {\bibinfo {volume}
			{48}},\ \bibinfo {pages} {102111} (\bibinfo {year} {2007})}\BibitemShut
	{NoStop}%
	\bibitem [{\citenamefont {Ge}\ \emph {et~al.}(2019)\citenamefont {Ge},
		\citenamefont {Tura},\ and\ \citenamefont {Cirac}}]{Ge_2019}%
	\BibitemOpen
	\bibfield  {author} {\bibinfo {author} {\bibfnamefont {Y.}~\bibnamefont
			{Ge}}, \bibinfo {author} {\bibfnamefont {J.}~\bibnamefont {Tura}},\ and\
		\bibinfo {author} {\bibfnamefont {J.~I.}\ \bibnamefont {Cirac}},\ }\bibfield
	{title} {\bibinfo {title} {Faster ground state preparation and high-precision
			ground energy estimation with fewer qubits},\ }\href
	{https://doi.org/10.1063/1.5027484} {\bibfield  {journal} {\bibinfo
			{journal} {Journal of Mathematical Physics}\ }\textbf {\bibinfo {volume}
			{60}},\ \bibinfo {pages} {022202} (\bibinfo {year} {2019})}\BibitemShut
	{NoStop}%
	\bibitem [{\citenamefont {Poulin}\ and\ \citenamefont
		{Wocjan}(2009)}]{Poulin_2009}%
	\BibitemOpen
	\bibfield  {author} {\bibinfo {author} {\bibfnamefont {D.}~\bibnamefont
			{Poulin}}\ and\ \bibinfo {author} {\bibfnamefont {P.}~\bibnamefont
			{Wocjan}},\ }\bibfield  {title} {\bibinfo {title} {Preparing ground states of
			quantum many-body systems on a quantum computer},\ }\href
	{https://doi.org/10.1103/physrevlett.102.130503} {\bibfield  {journal}
		{\bibinfo  {journal} {Physical Review Letters}\ }\textbf {\bibinfo {volume}
			{102}},\ \bibinfo {pages} {130503} (\bibinfo {year} {2009})}\BibitemShut
	{NoStop}%
	\bibitem [{\citenamefont {Lin}\ and\ \citenamefont {Tong}(2020)}]{Lin2020}%
	\BibitemOpen
	\bibfield  {author} {\bibinfo {author} {\bibfnamefont {L.}~\bibnamefont
			{Lin}}\ and\ \bibinfo {author} {\bibfnamefont {Y.}~\bibnamefont {Tong}},\
	}\bibfield  {title} {\bibinfo {title} {Near-optimal ground state
			preparation},\ }\href {https://doi.org/10.22331/q-2020-12-14-372} {\bibfield
		{journal} {\bibinfo  {journal} {Quantum}\ }\textbf {\bibinfo {volume} {4}},\
		\bibinfo {pages} {372} (\bibinfo {year} {2020})}\BibitemShut {NoStop}%
	\bibitem [{\citenamefont {Cubitt}\ and\ \citenamefont
		{Montanaro}(2016)}]{Cubitt_2016}%
	\BibitemOpen
	\bibfield  {author} {\bibinfo {author} {\bibfnamefont {T.}~\bibnamefont
			{Cubitt}}\ and\ \bibinfo {author} {\bibfnamefont {A.}~\bibnamefont
			{Montanaro}},\ }\bibfield  {title} {\bibinfo {title} {Complexity
			classification of local hamiltonian problems},\ }\href
	{https://doi.org/10.1137/140998287} {\bibfield  {journal} {\bibinfo
			{journal} {{SIAM} Journal on Computing}\ }\textbf {\bibinfo {volume} {45}},\
		\bibinfo {pages} {268} (\bibinfo {year} {2016})}\BibitemShut {NoStop}%
	\bibitem [{\citenamefont {Kempe}\ \emph {et~al.}(2006)\citenamefont {Kempe},
		\citenamefont {Kitaev},\ and\ \citenamefont {Regev}}]{Kempe_2006}%
	\BibitemOpen
	\bibfield  {author} {\bibinfo {author} {\bibfnamefont {J.}~\bibnamefont
			{Kempe}}, \bibinfo {author} {\bibfnamefont {A.}~\bibnamefont {Kitaev}},\ and\
		\bibinfo {author} {\bibfnamefont {O.}~\bibnamefont {Regev}},\ }\bibfield
	{title} {\bibinfo {title} {The complexity of the local hamiltonian problem},\
	}\href {https://doi.org/10.1137/s0097539704445226} {\bibfield  {journal}
		{\bibinfo  {journal} {{SIAM} Journal on Computing}\ }\textbf {\bibinfo
			{volume} {35}},\ \bibinfo {pages} {1070} (\bibinfo {year}
		{2006})}\BibitemShut {NoStop}%
	\bibitem [{\citenamefont {Golub}\ and\ \citenamefont {van~der
			Vorst}(2000)}]{Golub_2000}%
	\BibitemOpen
	\bibfield  {author} {\bibinfo {author} {\bibfnamefont {G.~H.}\ \bibnamefont
			{Golub}}\ and\ \bibinfo {author} {\bibfnamefont {H.~A.}\ \bibnamefont
			{van~der Vorst}},\ }\bibfield  {title} {\bibinfo {title} {Eigenvalue
			computation in the 20th century},\ }\href
	{https://doi.org/10.1016/s0377-0427(00)00413-1} {\bibfield  {journal}
		{\bibinfo  {journal} {Journal of Computational and Applied Mathematics}\
		}\textbf {\bibinfo {volume} {123}},\ \bibinfo {pages} {35} (\bibinfo {year}
		{2000})}\BibitemShut {NoStop}%
	\bibitem [{\citenamefont {Brassard}\ and\ \citenamefont
		{Hoyer}(1997)}]{Brassard1997}%
	\BibitemOpen
	\bibfield  {author} {\bibinfo {author} {\bibfnamefont {G.}~\bibnamefont
			{Brassard}}\ and\ \bibinfo {author} {\bibfnamefont {P.}~\bibnamefont
			{Hoyer}},\ }\bibfield  {title} {\bibinfo {title} {An exact quantum
			polynomial-time algorithm for simon{\textquotesingle}s problem},\ }in\ \href
	{https://doi.org/10.1109/istcs.1997.595153} {\emph {\bibinfo {booktitle}
			{Proceedings of the Fifth Israeli Symposium on Theory of Computing and
				Systems}}}\ (\bibinfo  {publisher} {{IEEE} Comput. Soc},\ \bibinfo {year}
	{1997})\BibitemShut {NoStop}%
	\bibitem [{\citenamefont {Grover}(1998)}]{Grover_1998}%
	\BibitemOpen
	\bibfield  {author} {\bibinfo {author} {\bibfnamefont {L.~K.}\ \bibnamefont
			{Grover}},\ }\bibfield  {title} {\bibinfo {title} {Quantum computers can
			search rapidly by using almost any transformation},\ }\href
	{https://doi.org/10.1103/physrevlett.80.4329} {\bibfield  {journal} {\bibinfo
			{journal} {Physical Review Letters}\ }\textbf {\bibinfo {volume} {80}},\
		\bibinfo {pages} {4329} (\bibinfo {year} {1998})}\BibitemShut {NoStop}%
	\bibitem [{\citenamefont {Yoder}\ \emph {et~al.}(2014)\citenamefont {Yoder},
		\citenamefont {Low},\ and\ \citenamefont {Chuang}}]{Yoder_2014}%
	\BibitemOpen
	\bibfield  {author} {\bibinfo {author} {\bibfnamefont {T.~J.}\ \bibnamefont
			{Yoder}}, \bibinfo {author} {\bibfnamefont {G.~H.}\ \bibnamefont {Low}},\
		and\ \bibinfo {author} {\bibfnamefont {I.~L.}\ \bibnamefont {Chuang}},\
	}\bibfield  {title} {\bibinfo {title} {Fixed-point quantum search with an
			optimal number of queries},\ }\href
	{https://doi.org/10.1103/physrevlett.113.210501} {\bibfield  {journal}
		{\bibinfo  {journal} {Physical Review Letters}\ }\textbf {\bibinfo {volume}
			{113}},\ \bibinfo {pages} {210501} (\bibinfo {year} {2014})}\BibitemShut
	{NoStop}%
	\bibitem [{\citenamefont {Dob{\v{s}}{\'{\i}}{\v{c}}ek}\ \emph
		{et~al.}(2007)\citenamefont {Dob{\v{s}}{\'{\i}}{\v{c}}ek}, \citenamefont
		{Johansson}, \citenamefont {Shumeiko},\ and\ \citenamefont
		{Wendin}}]{Dob_ek_2007}%
	\BibitemOpen
	\bibfield  {author} {\bibinfo {author} {\bibfnamefont {M.}~\bibnamefont
			{Dob{\v{s}}{\'{\i}}{\v{c}}ek}}, \bibinfo {author} {\bibfnamefont
			{G.}~\bibnamefont {Johansson}}, \bibinfo {author} {\bibfnamefont
			{V.}~\bibnamefont {Shumeiko}},\ and\ \bibinfo {author} {\bibfnamefont
			{G.}~\bibnamefont {Wendin}},\ }\bibfield  {title} {\bibinfo {title}
		{Arbitrary accuracy iterative quantum phase estimation algorithm using a
			single ancillary qubit: A two-qubit benchmark},\ }\href
	{https://doi.org/10.1103/physreva.76.030306} {\bibfield  {journal} {\bibinfo
			{journal} {Physical Review A}\ }\textbf {\bibinfo {volume} {76}},\ \bibinfo
		{pages} {030306} (\bibinfo {year} {2007})}\BibitemShut {NoStop}%
	\bibitem [{\citenamefont {Vedral}\ \emph {et~al.}(1996)\citenamefont {Vedral},
		\citenamefont {Barenco},\ and\ \citenamefont {Ekert}}]{Vedral_1996}%
	\BibitemOpen
	\bibfield  {author} {\bibinfo {author} {\bibfnamefont {V.}~\bibnamefont
			{Vedral}}, \bibinfo {author} {\bibfnamefont {A.}~\bibnamefont {Barenco}},\
		and\ \bibinfo {author} {\bibfnamefont {A.}~\bibnamefont {Ekert}},\ }\bibfield
	{title} {\bibinfo {title} {Quantum networks for elementary arithmetic
			operations},\ }\href {https://doi.org/10.1103/physreva.54.147} {\bibfield
		{journal} {\bibinfo  {journal} {Physical Review A}\ }\textbf {\bibinfo
			{volume} {54}},\ \bibinfo {pages} {147} (\bibinfo {year} {1996})}\BibitemShut
	{NoStop}%
	\bibitem [{\citenamefont {Garc{\'{\i}}a-P{\'{e}}rez}\ \emph
		{et~al.}(2020)\citenamefont {Garc{\'{\i}}a-P{\'{e}}rez}, \citenamefont
		{Rossi},\ and\ \citenamefont {Maniscalco}}]{Garc_a_P_rez_2020}%
	\BibitemOpen
	\bibfield  {author} {\bibinfo {author} {\bibfnamefont {G.}~\bibnamefont
			{Garc{\'{\i}}a-P{\'{e}}rez}}, \bibinfo {author} {\bibfnamefont {M.~A.~C.}\
			\bibnamefont {Rossi}},\ and\ \bibinfo {author} {\bibfnamefont
			{S.}~\bibnamefont {Maniscalco}},\ }\bibfield  {title} {\bibinfo {title}
		{{IBM} q experience as a versatile experimental testbed for simulating open
			quantum systems},\ }\href {https://doi.org/10.1038/s41534-019-0235-y}
	{\bibfield  {journal} {\bibinfo  {journal} {npj Quantum Information}\
		}\textbf {\bibinfo {volume} {6}},\ \bibinfo {pages} {1} (\bibinfo {year}
		{2020})}\BibitemShut {NoStop}%
	\bibitem [{\citenamefont {Berry}\ \emph
		{et~al.}(2015{\natexlab{a}})\citenamefont {Berry}, \citenamefont {Childs},
		\citenamefont {Cleve}, \citenamefont {Kothari},\ and\ \citenamefont
		{Somma}}]{Berry_2015}%
	\BibitemOpen
	\bibfield  {author} {\bibinfo {author} {\bibfnamefont {D.~W.}\ \bibnamefont
			{Berry}}, \bibinfo {author} {\bibfnamefont {A.~M.}\ \bibnamefont {Childs}},
		\bibinfo {author} {\bibfnamefont {R.}~\bibnamefont {Cleve}}, \bibinfo
		{author} {\bibfnamefont {R.}~\bibnamefont {Kothari}},\ and\ \bibinfo {author}
		{\bibfnamefont {R.~D.}\ \bibnamefont {Somma}},\ }\bibfield  {title} {\bibinfo
		{title} {Simulating hamiltonian dynamics with a truncated taylor series},\
	}\href {https://doi.org/10.1103/physrevlett.114.090502} {\bibfield  {journal}
		{\bibinfo  {journal} {Physical Review Letters}\ }\textbf {\bibinfo {volume}
			{114}},\ \bibinfo {pages} {090502} (\bibinfo {year}
		{2015}{\natexlab{a}})}\BibitemShut {NoStop}%
	\bibitem [{\citenamefont {Berry}\ \emph
		{et~al.}(2015{\natexlab{b}})\citenamefont {Berry}, \citenamefont {Childs},\
		and\ \citenamefont {Kothari}}]{Berry_2015_10}%
	\BibitemOpen
	\bibfield  {author} {\bibinfo {author} {\bibfnamefont {D.~W.}\ \bibnamefont
			{Berry}}, \bibinfo {author} {\bibfnamefont {A.~M.}\ \bibnamefont {Childs}},\
		and\ \bibinfo {author} {\bibfnamefont {R.}~\bibnamefont {Kothari}},\
	}\bibfield  {title} {\bibinfo {title} {Hamiltonian simulation with nearly
			optimal dependence on all parameters},\ }in\ \href
	{https://doi.org/10.1109/focs.2015.54} {\emph {\bibinfo {booktitle} {2015
				{IEEE} 56th Annual Symposium on Foundations of Computer Science}}}\ (\bibinfo
	{publisher} {{IEEE}},\ \bibinfo {year} {2015})\BibitemShut {NoStop}%
	\bibitem [{\citenamefont {Low}\ and\ \citenamefont {Chuang}(2017)}]{Low_2017}%
	\BibitemOpen
	\bibfield  {author} {\bibinfo {author} {\bibfnamefont {G.~H.}\ \bibnamefont
			{Low}}\ and\ \bibinfo {author} {\bibfnamefont {I.~L.}\ \bibnamefont
			{Chuang}},\ }\bibfield  {title} {\bibinfo {title} {Optimal hamiltonian
			simulation by quantum signal processing},\ }\href
	{https://doi.org/10.1103/physrevlett.118.010501} {\bibfield  {journal}
		{\bibinfo  {journal} {Physical Review Letters}\ }\textbf {\bibinfo {volume}
			{118}},\ \bibinfo {pages} {010501} (\bibinfo {year} {2017})}\BibitemShut
	{NoStop}%
	\bibitem [{\citenamefont {Low}\ and\ \citenamefont {Chuang}(2019)}]{Low_2019}%
	\BibitemOpen
	\bibfield  {author} {\bibinfo {author} {\bibfnamefont {G.~H.}\ \bibnamefont
			{Low}}\ and\ \bibinfo {author} {\bibfnamefont {I.~L.}\ \bibnamefont
			{Chuang}},\ }\bibfield  {title} {\bibinfo {title} {Hamiltonian simulation by
			qubitization},\ }\href {https://doi.org/10.22331/q-2019-07-12-163} {\bibfield
		{journal} {\bibinfo  {journal} {Quantum}\ }\textbf {\bibinfo {volume} {3}},\
		\bibinfo {pages} {163} (\bibinfo {year} {2019})}\BibitemShut {NoStop}%
	\bibitem [{\citenamefont {Killoran}\ \emph {et~al.}(2019)\citenamefont
		{Killoran}, \citenamefont {Bromley}, \citenamefont {Arrazola}, \citenamefont
		{Schuld}, \citenamefont {Quesada},\ and\ \citenamefont
		{Lloyd}}]{Killoran_2019}%
	\BibitemOpen
	\bibfield  {author} {\bibinfo {author} {\bibfnamefont {N.}~\bibnamefont
			{Killoran}}, \bibinfo {author} {\bibfnamefont {T.~R.}\ \bibnamefont
			{Bromley}}, \bibinfo {author} {\bibfnamefont {J.~M.}\ \bibnamefont
			{Arrazola}}, \bibinfo {author} {\bibfnamefont {M.}~\bibnamefont {Schuld}},
		\bibinfo {author} {\bibfnamefont {N.}~\bibnamefont {Quesada}},\ and\ \bibinfo
		{author} {\bibfnamefont {S.}~\bibnamefont {Lloyd}},\ }\bibfield  {title}
	{\bibinfo {title} {Continuous-variable quantum neural networks},\ }\href
	{https://doi.org/10.1103/physrevresearch.1.033063} {\bibfield  {journal}
		{\bibinfo  {journal} {Physical Review Research}\ }\textbf {\bibinfo {volume}
			{1}},\ \bibinfo {pages} {033063} (\bibinfo {year} {2019})}\BibitemShut
	{NoStop}%
	\bibitem [{\citenamefont {Zhao}\ and\ \citenamefont {Gao}(2021)}]{Zhao_2021}%
	\BibitemOpen
	\bibfield  {author} {\bibinfo {author} {\bibfnamefont {C.}~\bibnamefont
			{Zhao}}\ and\ \bibinfo {author} {\bibfnamefont {X.-S.}\ \bibnamefont {Gao}},\
	}\bibfield  {title} {\bibinfo {title} {{QDNN}: deep neural networks with
			quantum layers},\ }\href {https://doi.org/10.1007/s42484-021-00046-w}
	{\bibfield  {journal} {\bibinfo  {journal} {Quantum Machine Intelligence}\
		}\textbf {\bibinfo {volume} {3}},\ \bibinfo {pages} {15} (\bibinfo {year}
		{2021})}\BibitemShut {NoStop}%
	\bibitem [{\citenamefont {Panahiyan}\ and\ \citenamefont
		{Fritzsche}(2018)}]{Panahiyan_2018}%
	\BibitemOpen
	\bibfield  {author} {\bibinfo {author} {\bibfnamefont {S.}~\bibnamefont
			{Panahiyan}}\ and\ \bibinfo {author} {\bibfnamefont {S.}~\bibnamefont
			{Fritzsche}},\ }\bibfield  {title} {\bibinfo {title} {Controlling quantum
			random walk with a step-dependent coin},\ }\href
	{https://doi.org/10.1088/1367-2630/aad899} {\bibfield  {journal} {\bibinfo
			{journal} {New Journal of Physics}\ }\textbf {\bibinfo {volume} {20}},\
		\bibinfo {pages} {083028} (\bibinfo {year} {2018})}\BibitemShut {NoStop}%
	\bibitem [{\citenamefont {Lloyd}\ \emph {et~al.}(2014)\citenamefont {Lloyd},
		\citenamefont {Mohseni},\ and\ \citenamefont {Rebentrost}}]{Lloyd_2014}%
	\BibitemOpen
	\bibfield  {author} {\bibinfo {author} {\bibfnamefont {S.}~\bibnamefont
			{Lloyd}}, \bibinfo {author} {\bibfnamefont {M.}~\bibnamefont {Mohseni}},\
		and\ \bibinfo {author} {\bibfnamefont {P.}~\bibnamefont {Rebentrost}},\
	}\bibfield  {title} {\bibinfo {title} {Quantum principal component
			analysis},\ }\href {https://doi.org/10.1038/nphys3029} {\bibfield  {journal}
		{\bibinfo  {journal} {Nature Physics}\ }\textbf {\bibinfo {volume} {10}},\
		\bibinfo {pages} {631} (\bibinfo {year} {2014})}\BibitemShut {NoStop}%
	\bibitem [{\citenamefont {Aharonov}\ \emph {et~al.}(1993)\citenamefont
		{Aharonov}, \citenamefont {Davidovich},\ and\ \citenamefont
		{Zagury}}]{Aharonov_1993}%
	\BibitemOpen
	\bibfield  {author} {\bibinfo {author} {\bibfnamefont {Y.}~\bibnamefont
			{Aharonov}}, \bibinfo {author} {\bibfnamefont {L.}~\bibnamefont
			{Davidovich}},\ and\ \bibinfo {author} {\bibfnamefont {N.}~\bibnamefont
			{Zagury}},\ }\bibfield  {title} {\bibinfo {title} {Quantum random walks},\
	}\href {https://doi.org/10.1103/physreva.48.1687} {\bibfield  {journal}
		{\bibinfo  {journal} {Physical Review A}\ }\textbf {\bibinfo {volume} {48}},\
		\bibinfo {pages} {1687} (\bibinfo {year} {1993})}\BibitemShut {NoStop}%
	\bibitem [{\citenamefont {Venegas-Andraca}(2012)}]{Venegas_Andraca_2012}%
	\BibitemOpen
	\bibfield  {author} {\bibinfo {author} {\bibfnamefont {S.~E.}\ \bibnamefont
			{Venegas-Andraca}},\ }\bibfield  {title} {\bibinfo {title} {Quantum walks: a
			comprehensive review},\ }\href {https://doi.org/10.1007/s11128-012-0432-5}
	{\bibfield  {journal} {\bibinfo  {journal} {Quantum Information Processing}\
		}\textbf {\bibinfo {volume} {11}},\ \bibinfo {pages} {1015} (\bibinfo {year}
		{2012})}\BibitemShut {NoStop}%
\end{thebibliography}
\end{document}


\title{Supplemental Material of ``Quantum Heaviside Eigen Solver''}

	\author{Zheng-Zhi Sun}
	\affiliation{School of Physical Sciences, University of Chinese Academy of Sciences, P. O. Box 4588, Beijing 100049, China}

	\author{Gang Su}
	\email[Corresponding author. Email: ] {gsu@ucas.ac.cn}

	\affiliation{Kavli Institute for Theoretical Sciences, and CAS Center for Excellence in Topological Quantum Computation, University of Chinese Academy of Sciences, Beijing 100190, China}
	\affiliation{School of Physical Sciences, University of Chinese Academy of Sciences, P. O. Box 4588, Beijing 100049, China}

\maketitle

\section{Quantum Heaviside eigen solver}

\subsection{Quantum Heaviside and Dirac circuit}

Here we introduce the general form of QHC and QDC. To construct the circuit that can filter out all eigen states with eigen values larger than the trial threshold and preserve the rest states, the ideal QHC should satisfy
\begin{subequations}\label{eq-Heaviside-i}
	\begin{align}
		{{\bf{U}}_{Heaviside}}\left| {{E_j} < {\theta}} \right\rangle {\left| 0 \right\rangle ^{ \otimes K}}\left| 0 \right\rangle  = \left| {{E_j}} \right\rangle {\left| 0 \right\rangle ^{ \otimes K}}\left| 0 \right\rangle ,
	\end{align}
	\begin{align}
		{{\bf{U}}_{Heaviside}}\left| {{E_j} > {\theta}} \right\rangle {\left| 0 \right\rangle ^{ \otimes K}}\left| 0 \right\rangle  = \left| {{E_j}} \right\rangle {\left| {\varphi _j} \right\rangle}\left| 1 \right\rangle  .
	\end{align}
\end{subequations}
where $\left| {\varphi _j} \right\rangle $ represents the state that varies with different designs of QHC on $K$ auxiliary qubits. We do not care about $\left| {\varphi _j} \right\rangle $ as it does not influence the qualification of the states on physical qubits. The $\left| 0 \right\rangle $ and $\left| 1 \right\rangle $ in Eq. (\ref{eq-Heaviside-i}) are the states on the mark qubit, where $\left| 0 \right\rangle $ is entangled with the qualified states on physical qubits and $\left| 1 \right\rangle $ is entangled with other states.

However, this ideal QHC cannot be trivially realized in practice. So we define an approximation for the ideal QHC as
\begin{align}
	\label{eq-Heaviside-p}
	{{\bf{U}}_H}\left| {{E_j}} \right\rangle {\left| 0 \right\rangle ^{ \otimes K}}\left| 0 \right\rangle  = {\alpha _j}\left| {{E_j}} \right\rangle {\left| 0 \right\rangle ^{ \otimes K}}\left| 0 \right\rangle +   {\beta _j}\left| {{E_j} } \right\rangle \left| {{\varphi _j}} \right\rangle \left| 1 \right\rangle ,
\end{align}
where ${\left| {{\alpha _j}} \right|^2} + {\left| {{\beta _j}} \right|^2} = 1$. This definition is the same as Eq. (\ref{eq-Heaviside-i}) when ${\alpha _j}$ is a Heaviside function such that ${\alpha _j} = u\left(\theta - {{E_j} } \right)$, where $u$ is the classical unit step function.

Similarly, the qualified states for the QDC are the states $\left| {{E_g}} \right\rangle $ corresponding to a given eigen value ${{E_g}}$. To achieve the qualification of the state $\left| {{E_g}} \right\rangle $, the ideal QDC should satisfy the following conditions
\begin{subequations}\label{eq-Dirac-i}
	\begin{align}
		{{\bf{U}}_{Dirac}}\left| {{E_g}} \right\rangle {\left| 0 \right\rangle ^{ \otimes K}}\left| 0 \right\rangle  = \left| {{E_g}} \right\rangle {\left| 0 \right\rangle ^{ \otimes K}}\left| 0 \right\rangle ,
	\end{align}
	\begin{align}
		{{\bf{U}}_{Dirac}}\left| {{E_{j \ne g}}} \right\rangle {\left| 0 \right\rangle ^{ \otimes K}}\left| 0 \right\rangle  = \left| {{E_{j \ne g}}} \right\rangle {\left| {\phi _j} \right\rangle}\left| 1 \right\rangle ,
	\end{align}
\end{subequations}
This ideal QDC cannot be trivially realized in practice. So we define an approximation for the ideal QDC as
\begin{align}
	\label{eq-Dirac-p}
	{{\bf{U}}_D}\left| {{E_j}} \right\rangle {\left| 0 \right\rangle ^{ \otimes K}}\left| 0 \right\rangle  = {\gamma _j}\left| {{E_j}} \right\rangle {\left| 0 \right\rangle ^{ \otimes K}} \left| 0 \right\rangle  + {\rm{ }}{\eta _j}\left| {{E_j}} \right\rangle \left| {\phi _j} \right\rangle \left| 1 \right\rangle  ,
\end{align}
where ${\left| {{\gamma _j}} \right|^2} + {\left| {{\eta _j}} \right|^2} = 1$. This definition is the same as Eq. (\ref{eq-Dirac-i}) when ${\gamma _j}$ equals to $1$ instead of $0$ only if ${E_j} = {E_g}$, which is an analog to the Dirac function.

\subsection{Preliminaries of amplitude amplification algorithm}

Now we introduce the amplitude amplification algorithm used to solve the Hamiltonian matrix. Suppose that we have an initialization operator ${{\bf{U}}_I}$ on $N$ qubits which achieves $\left| {{\psi _r}} \right\rangle {\rm{ = }}{{\bf{U}}_I}{\left| 0 \right\rangle ^{ \otimes N}}$ and a projector ${\bf{P}} = \left| {{E_q}} \right\rangle \left\langle {{E_q}} \right|$ to the qualified state $\left| {{E_q}} \right\rangle $. Grover search algorithm \cite{Grover_1997} can project $\left| {{\psi _r}} \right\rangle $ onto the image of ${\bf{P}}$ with a small correction by performing $O\left( {{1 \over {\sqrt p }}} \right)$ iterations of ${\bf{I}} - 2{\bf{P}}$ and ${\bf{I}} - 2\left| {{\psi _r}} \right\rangle \left\langle {{\psi _r}} \right|$, where ${\left\| {{\bf{P}}\left| {{\psi _r}} \right\rangle } \right\|^2} = p$. The amplitude amplification algorithm performs a similar process in the case of multiple qualified states. To achieve the quadratic speedup, the initialization operator ${{\bf{U}}_I}$ should satisfy that the fidelity between qualified state and initialized state is no less than $O\left( {{1 \over \chi }} \right)$, which is a trivial task since even for a random initialization operator it has a high probability of feasibility \cite{Poulin_2009}. The main difficulty to solve the eigen problem of Hamiltonian is to design the projector to the qualified state, in particular when we do not have prior knowledge of the eigen values and eigen states.

After the construction of this projector using the QDC or QHC, the unitary Grover search circuit is recorded as
\begin{align}
	\label{eq-grover}
	{\bf{G}}\left( {{{\bf{U}}_I},{\bf{P}}} \right) = {\left[ {\left( {{\bf{I}} - 2{\bf{P}}} \right)\left( {{\bf{I}} - 2\left| {{\psi _r}} \right\rangle \left\langle {{\psi _r}} \right|} \right)} \right]^{O\left( {{1 \over {\sqrt p }}} \right)}}.
\end{align}
This circuit can amplify the amplitude of $\left| {{E_q}} \right\rangle $ to $O\left( 1 \right)$ starting with the randomly initialized state ${\left| {{\psi _r}} \right\rangle }$, which is given by
\begin{align}
	\label{eq-grover-p}
	\left| {{\bf{PG}}\left( {{{\bf{U}}_I},{\bf{P}}} \right)\left| {{\psi _r}} \right\rangle } \right| = O\left( 1 \right).
\end{align}
When solving the eigen problem using the amplitude amplification algorithm, the value of $p$ is generally not known in advance. This causes the soufflé problem where the Grover search algorithm is seen as a ``quantum oven'' \cite{Brassard_1997}. The soufflé problem is that the amplitude of the qualified state drops to zeros if you open the oven too early and the amplitude starts shrinking if using too many iterations. Luckily we can use either the full-blown quantum counting \cite{Boyer_1998, Brassard_1998} or a trial-and-error scheme where iterates are applied by an exponentially increasing number of times \cite{Boyer_1998, Brassard2002} without losing the quadratic speedup from quantum search algorithm. A more elegant method is the fixed-point search \cite{Yoder_2014}, which is an improved version of the amplitude amplification algorithm to avoid the ``over cooking'' problem. We adopt this fixed-point search as the amplitude amplification method here and record it as ${{\bf{F}}\left( {{{\bf{U}}_I},{\bf{P}}} \right)}$.

\subsection{Constructing projectors to qualified states}

We can construct the projector of QDC by combining ${\bf{U}}_D$ and a projector to ${\left| 0 \right\rangle ^{ \otimes K}}\left| 0 \right\rangle $ on the $K$ auxiliary qubits and the mark qubit as
\begin{align}
	\label{eq-Dirac-projector}
	{{\bf{P}}_D} = \left[ {{{\bf{I}}^{ \otimes N}} \otimes {{\left( {\left| 0 \right\rangle \left\langle 0 \right|} \right)}^{ \otimes K + 1}}} \right]{{\bf{U}}_D}.
\end{align}
Here ${\bf{U}}_D$ is the practical quantum Dirac circuit, ${{{\bf{I}}^{ \otimes  {N} }}}$ is the identity operator on the physical qubits, and ${\left( {\left| 0 \right\rangle \left\langle 0 \right|} \right)^{ \otimes K + 1}}$ is the projector to ${\left| 0 \right\rangle ^{ \otimes K}}\left| 0 \right\rangle $ on the $K$ auxiliary qubits and the mark qubit.
When the state is in the subspace where the auxiliary qubits and the mark qubit are fixed to ${\left| 0 \right\rangle ^{ \otimes K}}\left| 0 \right\rangle $, the projector ${{\bf{P}}_D}$ achieves the result of the projector $\left| {{E_g}} \right\rangle \left\langle {{E_g}} \right|$ on physical qubits. This can be expressed as
\begin{align}
	\label{eq-Dirac-projector-e}
	{{\bf{P}}_D}\left| {{E_j}} \right\rangle {\left| 0 \right\rangle ^{ \otimes K}}\left| 0 \right\rangle  = {\gamma _j}\left| {{E_j}} \right\rangle {\left| 0 \right\rangle ^{ \otimes K}}\left| 0 \right\rangle  .
\end{align}
The projector can be described as firstly entangling the target state $\left| {{E_g}} \right\rangle $ on the physical qubits with  ${\left| 0 \right\rangle }$ on the mark qubit and then projecting the mark qubit with $\left| 0 \right\rangle \left\langle 0 \right|$.

Similarly, the corresponding projector ${{\bf{P}}_H}$ of QHC can be constructed by combining ${{\bf{U}}_{H}}$ with a projector to ${\left| 0 \right\rangle ^{ \otimes K}}\left| 0 \right\rangle $ on the $K$ auxiliary qubits and the mark qubit, which gives
\begin{equation}
	\label{eq-Heaviside-projector}
	{{\bf{P}}_H} = \left[ {{{\bf{I}}^{ \otimes N}} \otimes {{\left( {\left| 0 \right\rangle \left\langle 0 \right|} \right)}^{ \otimes K + 1}}} \right]{{\bf{U}}_H}.
\end{equation}

\subsection{Quantum selector and quantum judge for eigen problem}

After constructing the projector in amplitude amplification algorithm using the quantum Heaviside circuit as given in Eq. (\ref{eq-Heaviside-projector}), a quantum judge can be obtained. The quantum judge projects a randomly initialized state onto the image of states whose energies are lower than the given threshold $\theta$ with high probability. Then an extra quantum Heaviside circuit is performed to mark the qualified states with exponentially small error. Measuring the mark qubit one can obtain whether there is any qualified state to the quantum Heaviside circuit of the given Hamiltonian matrix and threshold. More specifically, if the probability of the measurement result of the mark qubit being $\left| 0 \right\rangle $ is exponentially small, all eigen values of given Hamiltonian matrix are larger than $\theta$. The minimum eigen value of $\bf{H}$ with an error smaller than $\varepsilon $ can be obtained by performing $O\left( {\log {1 \over \varepsilon }} \right)$ times of quantum judge according to dichotomy. The number of calling quantum circuit to obtain the ground state energy of $\bf{H}$ is exponentially smaller than those in Ref. \cite{Poulin_2009} and Ref. \cite{Ge_2019}, while the cost of quantum circuits is similar. These two works use analogs of quantum Dirac circuit for sweeping or quantum sweeping the ground state energy for $O\left( {{1 \over \varepsilon }} \right)$ and $O\left( {\sqrt {{1 \over \varepsilon }} } \right)$ times, respectively, in the absence of quantum Heaviside circuit.

As the eigen values can be calculated by combining quantum judge with dichotomy, the quantum Dirac circuit can be used to find the eigen state of a given eigen value. The amplitude amplification algorithm whose projector is constructed by quantum Dirac circuit can be regarded as a quantum selector. Similar to quantum judge, the quantum selector projects a randomly initialized state onto the image of states corresponding to the given eigen value. Then an extra quantum Dirac circuit is performed to mark the qualified states with exponentially small error. When this process is done, the state $\left| 0 \right\rangle $ on the mark qubit is entangled with the qualified states on the physical qubits. A joint measurement of the mark qubit and physical qubits can be performed to make tomography or obtain the observations of qualified states. The quantum selector does not distinguish the degenerate states since it changes the mark qubit based on the eigen values of input states on the physical qubits. When the qualified states are degenerate (multiple), the state on the physical qubits entangled with  $\left| 0 \right\rangle $ on the mark qubit is a linear superposition of all qualified states. The superposition coefficients depend on the proportion of different qualified states in the initial state $\left| {{\psi _r}} \right\rangle $, which are usually unknown.

\section{Solve eigen states with quantum selector}

\subsection{Restrictions on quantum Dirac circuit}

Now we give the restrictions on the quantum Dirac to solve a $\chi $-dimensional Hamiltonian matrix on $N$ physical qubits. In the case of quantum Dirac circuit, the quantum selector is an amplitude amplification circuit whose projector to the good subspace is ${{\bf{P}}_D}$. The output state of this quantum selector on physical qubits where the mark qubit is measured to be $\left| 0 \right\rangle $ is
\begin{align}
	\label{eq-output1}
	{{{{\bf{P}}_D}\left| {{\psi _r}} \right\rangle } \over {\left\| {{{\bf{P}}_D}\left| {{\psi _r}} \right\rangle } \right\|}} = {{\sum\limits_j {{\lambda _j}{\gamma _j}\left| {{E_j}} \right\rangle } } \over {\left\| {\sum\limits_j {{\lambda _j}{\gamma _j}\left| {{E_j}} \right\rangle } } \right\|}}  ,
\end{align}
where  $\left\{ {{\lambda _j}} \right\}$ are the expansion coefficients of initial state under energy representations. These coefficients are given by the initialization circuit
\begin{align}
	\label{eq-initial-circuit}
	\left| {{\psi _r}} \right\rangle  = {{\bf{U}}_I}{\left| 0 \right\rangle ^{ \otimes N}} = \sum\limits_j {{\lambda _j}\left| {{E_j}} \right\rangle }   .
\end{align}

Without losing generality, we assume that the qualified state is ${\left| {{E_g}} \right\rangle }$. And we also suppose that a suitable initialization operator is constructed to satisfy that ${\lambda _g} = O\left( {{1 \over {\sqrt \chi  }}} \right)$. The eigen state corresponding to ${E_g}$ is assumed to be non-degenerate for the convenience of explanation. This algorithm can be trivially generalized to the degenerate case. The successful implementation of using quantum selector to obtain an $\varepsilon $-close state to $\left| {{E_g}} \right\rangle $ has two requirements that come from the amplitude amplification algorithm \cite{Grover_1998}. The first requirement is that the amplitude amplification circuit can be constructed by no more than $O\left( {{1 \over {\sqrt \chi  }}} \right)$ initialization circuits ${{\bf{U}}_I}$ and projector circuits ${{\bf{P}}_D}$, which is equivalent to that $\left\| {{{\bf{P}}_D}\left| {{\psi _r}} \right\rangle } \right\| \ge O\left( {{1 \over {\sqrt \chi  }}} \right)$. The second requirement is that the error between the output state of this quantum selector and ${\left| {{E_g}} \right\rangle }$ should be less than $O\left( \varepsilon  \right)$, which can be expressed as $\left\| {{{{{\bf{P}}_D}\left| {{\psi _r}} \right\rangle } \over {\left\| {{{\bf{P}}_D}\left| {{\psi _r}} \right\rangle } \right\|}} - \left| {{E_g}} \right\rangle } \right\| \le O\left( \varepsilon  \right)$.

To meet these two requirements, we restrain that $\left\{ {{\gamma _j}} \right\}$ in Eq. (\ref{eq-Dirac-p}) satisfy
\begin{equation}\label{eq-restrain-alpha}
	\left| {{\gamma _j}} \right|
	\begin{cases}
		\ge {1 \over 2} & j = g, \\
		\le O\left( {{\varepsilon  \over {\sqrt \chi  }}} \right) & j \ne g,
	\end{cases}
\end{equation}
which is a sufficient condition for these two requirements when designing the practical quantum Dirac circuit of Eq. (\ref{eq-Dirac-p}). The proof of sufficiency for the first requirement is that
\begin{equation}
	\label{eq-Dirac-proof1}
	\left\| {{{\bf{P}}_D}\left| {{\psi _r}} \right\rangle } \right\| \ge \left\| {{\lambda _g}{\gamma _g}\left| {{E_g}} \right\rangle } \right\| = \left| {{\lambda _g}{\gamma _g}} \right| = O\left( {{1 \over {\sqrt \chi  }}} \right).
\end{equation}
Meanwhile, we notice that
\begin{subequations}\label{eq-Dirac-subproof}
	\begin{align}
		\sum\limits_j {{{\left| {{\lambda _j}{\gamma _j}} \right|}^2}}  \ge {\left| {{\lambda _g}{\gamma _g}} \right|^2} = O\left( {{1 \over \chi }} \right) ,
	\end{align}
	\begin{align}
		\sum\limits_{i \ne g} {{{\left| {{\lambda _j}{\gamma _j}} \right|}^2}}  \le O\left( {{{{\varepsilon ^2}} \over \chi }} \right)\sum\limits_{i \ne g} {{{\left| {{\lambda _j}} \right|}^2}}  = O\left( {{{{\varepsilon ^2}} \over \chi }} \right)  .
	\end{align}
\end{subequations}
The proof of sufficiency for the second requirement is that
\begin{align}
	\label{eq-Dirac-proof2}
	{{\left| {\left\langle {{E_g}} \right|{{\bf{P}}_D}\left| {{\psi _r}} \right\rangle } \right|} \over {\left\| {{{\bf{P}}_D}\left| {{\psi _r}} \right\rangle } \right\|}} =
	{{\left| {{\lambda _g}{\gamma _g}} \right|} \over {\sqrt {\sum\limits_j {{{\left| {{\lambda _j}{\gamma _j}} \right|}^2}} } }} \\ \notag
	= \sqrt {1 - {{\sum\limits_{i \ne g} {{{\left| {{\lambda _j}{\gamma _j}} \right|}^2}} } \over {\sum\limits_j {{{\left| {{\lambda _j}{\gamma _j}} \right|}^2}} }}}   \\ \notag
	\ge \sqrt {1 - O\left( {{\varepsilon ^2}} \right)} ,
\end{align}
and $\left\| {{{{{\bf{P}}_D}\left| {{\psi _r}} \right\rangle } \over {\left\| {{{\bf{P}}_D}\left| {{\psi _r}} \right\rangle } \right\|}} - \left| {{E_q}} \right\rangle } \right\| \le O\left( \varepsilon  \right)$ can be directly derived.

\subsection{Realization of quantum coin toss}
Here we give the quantum circuit to flip one quantum coin with the amplitude of $\cos \left( {{E_j} - {E_g}} \right)$ when the state on the physical qubits is $\left| {{E_j}} \right\rangle $, where ${{E_g}}$ is the given eigen value. Since we can shift and zoom the Hamiltonian matrix $\bf{H}$, we assume that ${{E_g} = 0}$ and all eigen values of $\bf{H}$ are in the range of $\left( { - {\pi  \over 2},{\pi  \over 2}} \right)$ without losing generality. This flipping operator of quantum coin is designed as
\begin{align}
	\label{eq-uf-one}
	{{\bf{U}}_c} = {\bf{B}}{{\bf{U}}_e}{\bf{B}},
\end{align}
where ${\bf{B}}$ is the quantum Hadamard gate which maps basis state $\left| 0 \right\rangle $ to ${{\left| 0 \right\rangle  + \left| 1 \right\rangle } \over {\sqrt 2 }}$ and maps $\left| 1 \right\rangle $ to ${{\left| 0 \right\rangle  - \left| 1 \right\rangle } \over {\sqrt 2 }}$ \cite{Nielsen_2009}, and $\bf{U}_e$ is a time evolution operator controlled by the quantum coin
\begin{align}
	\label{eq-u-evolute}
	{{\bf{U}}_e} = {e^{i{\bf{H}}}}\left| 0 \right\rangle \left\langle 0 \right| + {e^{ - i{\bf{H}}}}\left| 1 \right\rangle \left\langle 1 \right|.
\end{align}
${e^{ - i{\bf{H}}}}$ is the unit time evolution of Hamiltonian matrix $\bf{H}$ and ${e^{ i{\bf{H}}}}$ is the inverse of it. The illustration of this design is shown in Fig. \ref{fig-dirac-1}(c).

\subsection{Using multiple quantum coins to design quantum Dirac circuit}

\begin{figure}[htb]
	\centering
	\includegraphics[width=0.9\linewidth]{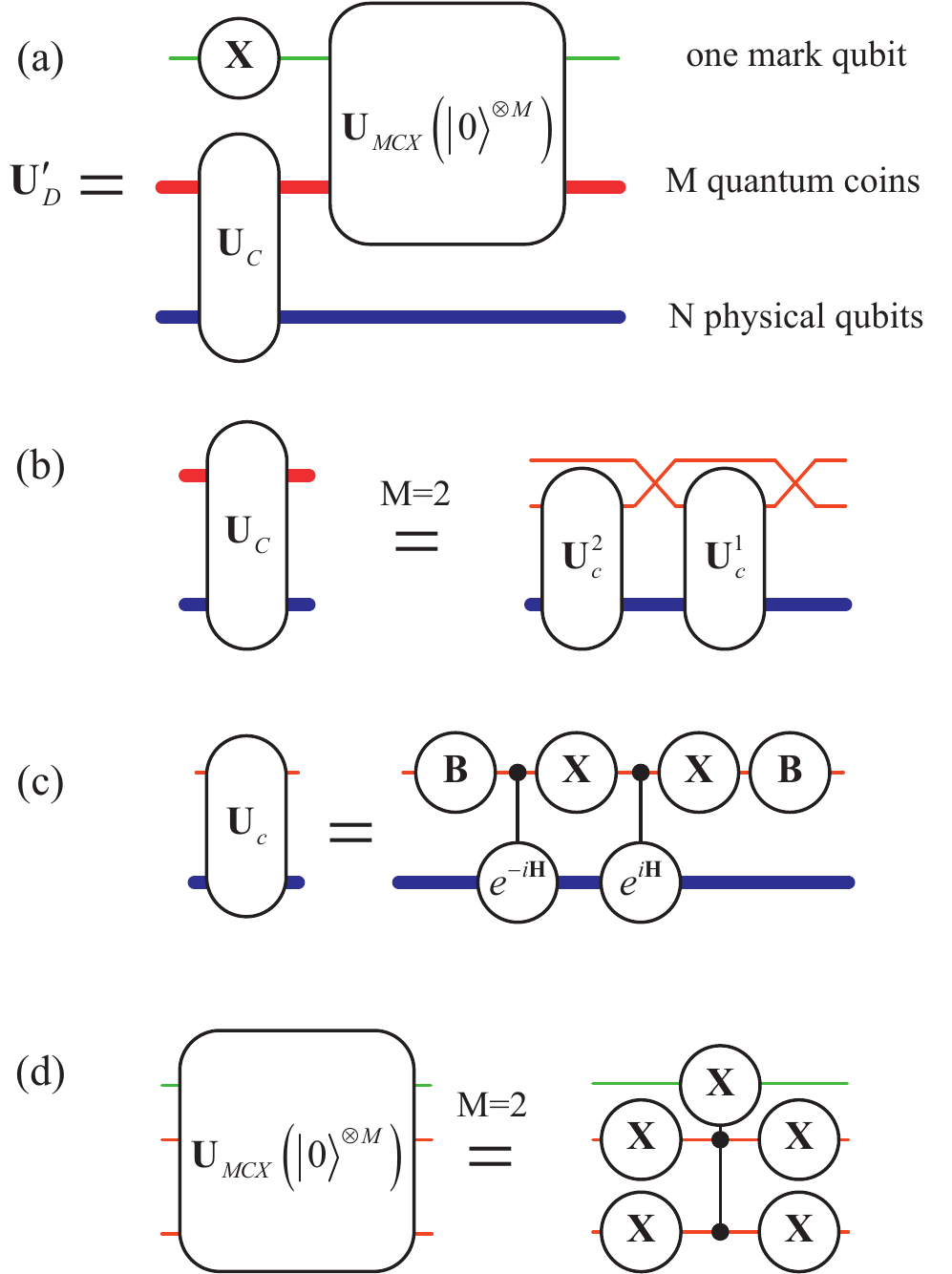}
	\caption{\label{fig-dirac-1}(a) guide to construct the primary quantum Dirac circuit of Eq. (\ref{eq-Dirac-1}). (a) The primary design of the quantum Dirac circuit of Eq. (\ref{eq-Dirac-1}). (b) The circuit that flips $M$ quantum coins respectively as given in Eq. (\ref{eq-ufM-result}). Here we record the top quantum coin as the first quantum coin. (c) The flipping operator of one quantum coin in Eq. (\ref{eq-uf-one}). (d) The multi-controlled NOT gate that performs NOT gate on the mark qubit when the state on $M$ quantum coins is ${{{\left| 0 \right\rangle }^{ \otimes M}}}$.}
\end{figure}

Here we consider the operator ${{\bf{U}}_C}$ that flips $M$ quantum coins
\begin{align}
	\label{eq-ufM-result}
	{{\bf{U}}_C}\left| {{E_j}} \right\rangle & {\left| 0 \right\rangle ^{ \otimes M}} = \left( {\prod\limits_m^M {{\bf{U}}_c^m} } \right)\left| {{E_j}} \right\rangle {\left| 0 \right\rangle ^{ \otimes M}} \notag \\
	& = \left| {{E_j}} \right\rangle {\left[ {{{\cos }}\left( {{E_j}} \right)\left| 0 \right\rangle  + i\sin \left( {{E_j}} \right)\left| 1 \right\rangle } \right]^{ \otimes M}} .
\end{align}
${{\bf{U}}_c^m}$ is an operator on $N$ physical qubits and $M$ quantum coins, which is constructed by flipping operator ${{\bf{U}}_c}$ of the $m$-th quantum coin and an identity operator on the rest $M-1$ quantum coins, as shown in Fig. \ref{fig-dirac-1}(b). We call this kind of operator composed by the direct product of a unitary operator $\bf{U}$ and an identity operator on the rest qubits of the whole circuit as the extension of $\bf{U}$ below. The probability of all quantum coins being flipped is ${{{\cos }^{2M}}\left( {{E_j}} \right)}$, which is a good approximation to the analog of Dirac function $\delta \left( {{E_j}} \right)$ for large $M$. The practical quantum Dirac circuit in the form of Eq. (\ref{eq-Dirac-p}) can be constructed by
\begin{align}
	\label{eq-Dirac-1}
	{{\bf{U'}}_D} = \left[ {{{\bf{I}}^{ \otimes N}} \otimes {{\bf{U}}_{MCX}}\left( {{{\left| 0 \right\rangle }^{ \otimes M}}} \right)} \right]\left( {{{\bf{U}}_C} \otimes {\bf{I}}} \right)\left( {{{\bf{I}}^{ \otimes N + M}} \otimes \bf{X}} \right) ,
\end{align}
which is shown in Fig. \ref{fig-dirac-1}(a). ${{{\bf{U}}_{MCX}}\left( {{{\left| 0 \right\rangle }^{ \otimes M}}} \right)}$ is a multi-controlled NOT gate that performs NOT gate on the mark qubit when the state on $M$ quantum coins is ${{{\left| 0 \right\rangle }^{ \otimes M}}}$ as shown in Fig. \ref{fig-dirac-1}(d). The operator ${{\bf{U}}_C}$ is the combination of $M$ flipping operator acting on the physical qubits and $M$ quantum coins separately as shown in Fig. \ref{fig-dirac-1}(b). ${\bf{X}}$ represents for the quantum NOT gate.

\subsection{Mathematical analysis on the number of quantum coins}

With the practical design of Eq. (\ref{eq-Dirac-1}), we can obtain from Eq. (\ref{eq-Dirac-p}) that ${\gamma _j} = {\cos ^M}\left( {{E_j}} \right)$. The restraint of ${\gamma _j}$ in Eq. (\ref{eq-restrain-alpha}) becomes the restraint of $M$ and the error bound ${{\varepsilon _0}}$ of the given energy ${E_g}$
\begin{subequations}\label{eq-reatrain-M}
	\begin{align}
		{\cos ^M}\left( {{\varepsilon _0}} \right) \ge {1 \over 2},
	\end{align}
	\begin{align}
		{\cos ^M}\left( {\left| {\Delta  - {\varepsilon _0}} \right|} \right) < O\left( {{\varepsilon  \over {\sqrt \chi  }}} \right) ,
	\end{align}
\end{subequations}
where $\varepsilon $ is the error bound of the output state, $\chi $ is the dimension of Hamiltonian matrix $\bf{H}$, and $\Delta $ is the gap between ${E_g}$ and its closest eigen value. Then we can derive that to obtain the $\varepsilon$-close eigen state corresponding to the given eigen value ${E_g}$, the minimum number of quantum coins is 
\begin{align}
	\label{eq-M}
	M = O\left( {{1 \over {{\Delta ^2}}}\left( {N + \log {1 \over \varepsilon }} \right)} \right),
\end{align}
and the error bound of ${E_g}$ is ${\varepsilon _0} = O\left( \Delta  \right)$.

\subsection{Quantum Dirac circuit improved by the freezing operator}

\begin{figure}[htb]
	\centering
	\includegraphics[width=1\linewidth]{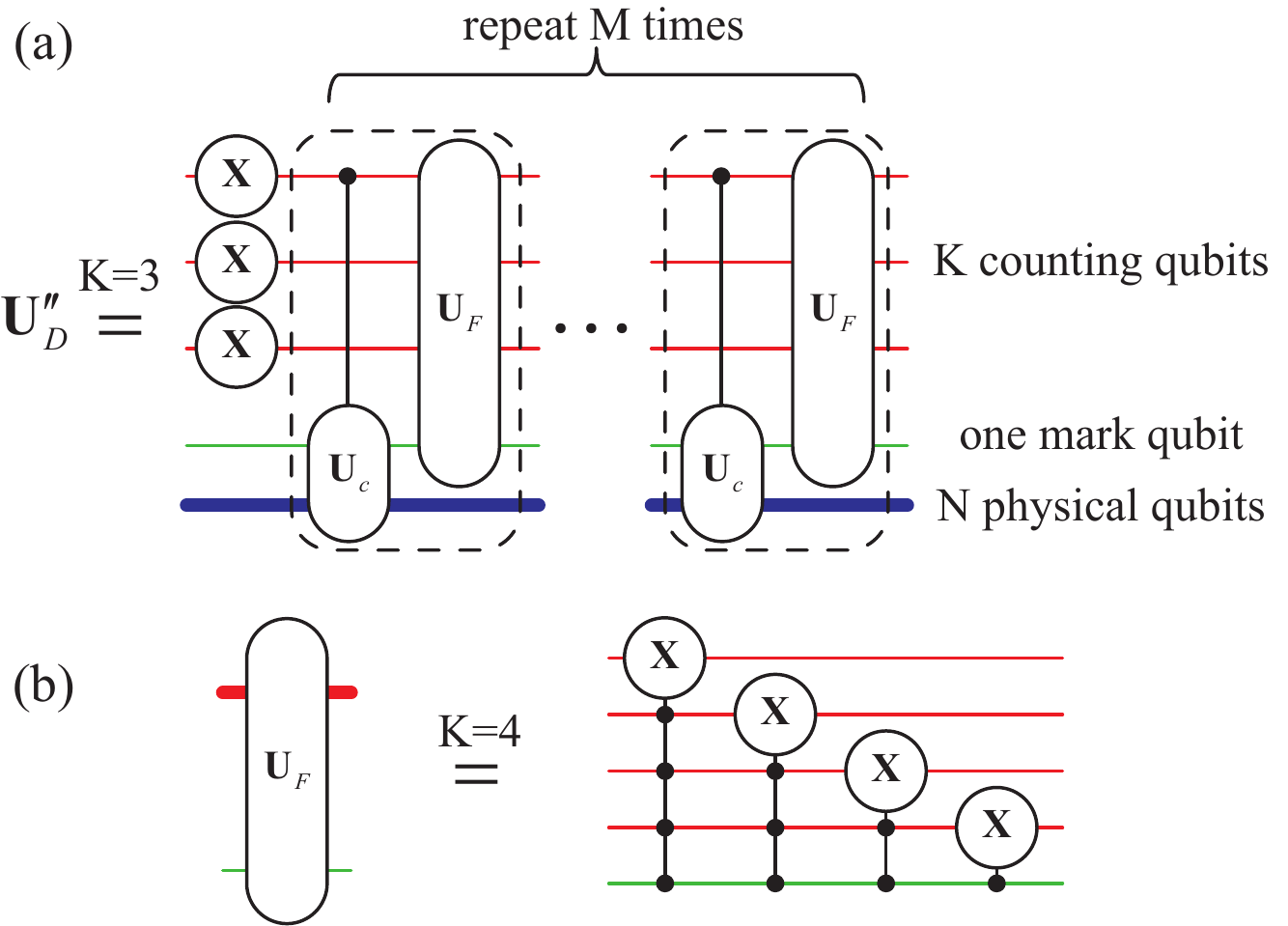}
	\caption{\label{fig-dirac-2}(a) guide to construct the quantum Dirac circuit of Eq. (\ref{eq-Dirac-2}). (a) The design of quantum Dirac circuit of Eq. (\ref{eq-Dirac-2}) used for the quantum selector. Here we record the top counting qubit as the first (the highest) qubit and so do the following circuits. (b) The circuit of freezing operator.}
\end{figure}

The new design of practical quantum Dirac circuit is shown in Fig. \ref{fig-dirac-2} and can be written as
\begin{align}
	\label{eq-Dirac-2}
	{{\bf{U''}}_D} = {\left[ {\left( {{{\bf{I}}^{ \otimes N}} \otimes {{\bf{U}}_F}} \right)C{{\bf{U}}_c}} \right]^M}\left( {{{\bf{I}}^{ \otimes N + 1}} \otimes {{\bf{X}}^{ \otimes K}}} \right),
\end{align}
where ${C{{\bf{U}}_c}}$ is the flipping operator controlled by the mark qubit and combined with an identity operator on the rest $K - 1$ qubits. This method shows how to construct the practical quantum Dirac circuit in the form of Eq. (\ref{eq-Dirac-p}) with $K = O\left( {\log M} \right)$ auxiliary qubits (counting qubits).

\subsection{Construction and gate complexity of quantum selector}

Using Eq. (\ref{eq-Dirac-2}) as the practical quantum Dirac circuit, the quantum selector can be expressed as
\begin{align}
	\label{eq-selector}
	{\bf{S}} = {{\bf{P''}}_D}{\bf{F}}\left[ {{{\bf{U}}_I},{{{\bf{P''}}}_D}} \right],
\end{align}
where ${{\bf{P''}}_D}$ is the projector to the good subspace constructed by Eq. (\ref{eq-Dirac-2}) and Eq. (\ref{eq-Dirac-projector}), which reads
\begin{align}
	\label{eq-D-projector}
	{{\bf{P''}}_D} = \left[ {{{\bf{I}}^{ \otimes N + K}} \otimes \left( {\left| 0 \right\rangle \left\langle 0 \right|} \right)} \right]{\bf{U''}_D}.
\end{align}
Note that this projector is not a unitary operator as it contains ${\left| 0 \right\rangle \left\langle 0 \right|}$ which is achieved by measurements. This quantum selector can output an $\varepsilon$-close eigen state corresponding to the given eigen value $E_g$. Since the quantum selector is a quantum amplitude amplification algorithm, it requires $O\left( {{1 \over {\sqrt \chi  }}} \right)$ iterations of the initialization operator ${{{\bf{U}}_I}}$ and the projector ${{{{\bf{P''}}}_D}}$.

Now we analyze the gate complexity of the quantum selector. The basic ingredients of the quantum selector include the initialization operator ${{\bf{U}}_I}$ defined by Eq. (\ref{eq-initial-circuit}), the controlled Hamiltonian evolution operator ${{\bf{U}}_e}$ defined by Eq. (\ref{eq-u-evolute}), the multi-controlled NOT gate, and some basic quantum gates. There are several methods for the controlled Hamiltonian evolution whose gate complexities scale differently with the evolution time, the error bound and the properties of the given Hamiltonian matrix \cite{Berry_2015, Berry_2015_10, Low_2017, Low_2019}. Here we record the gate complexity for Hamiltonian simulation in unit time as $\Lambda $ instead of analyzing the difference of gate complexities resulted from different Hamiltonian simulation methods. Similarly, we record the gate complexity of initialization circuit as $\Phi $ from which one can obtain the state whose fidelity with the qualified state is no less than $O\left( {{1 \over \chi }} \right)$.

From Fig. \ref{fig-dirac-1}(c) we can see that the gate complexity of ${{\bf{U}}_c}$ is $O\left( \Lambda  \right)$. The gate complexity of ${{\bf{U}}_F}$ is $O\left[ {poly\left( K \right)} \right]$ as shown in Fig. \ref{fig-dirac-2}(b), and the gate complexity of multi-controlled NOT gate on $K+1$ qubits is also $O\left[ {poly\left( K \right)} \right]$ \cite{Nielsen_2009}. Then the gate complexity of the quantum Dirac circuit of Eq. (\ref{eq-Dirac-2}) is $O\left( {M\left[ {\Lambda  + poly\left( K \right)} \right]} \right)$. We use $\tilde O$ to denote the complexity up to poly logarithmic factors in $N$, ${1 \over \Delta }$, ${\log {1 \over \varepsilon }}$, and $\Lambda $. The gate complexity of the quantum Dirac circuit of Eq. (\ref{eq-Dirac-2}) can thus be rewritten as $\tilde O\left( {\Lambda {{N + \log {1 \over \varepsilon }} \over {{\Delta ^2}}}} \right)$. Since the quantum selector is the Grove search of Eq. (\ref{eq-grover}) (or the fixed-point search with the same complexity under big-$O$ notation) using ${{\bf{U}}_I}$ and ${{\bf{P}}_D}$ of Eq. (\ref{eq-D-projector}), its gate complexity is
\begin{align}
	\label{eq-gate-dirac}
	\tilde O\left( {\Lambda {{N + \log {1 \over \varepsilon }} \over {\sqrt \chi  {\Delta ^2}}} + {\Phi  \over {\sqrt \chi  }}} \right).
\end{align}

\section{Solve eigen values with quantum judge}

\subsection{Restrictions on quantum Heaviside circuit}

The quantum judge is an amplitude amplification circuit whose projector to the good subspace is ${{\bf{P}}_H}$. Then the output state of this quantum judge on physical qubits where the mark qubit is measured to be $\left| 0 \right\rangle $ is
\begin{align}
	\label{eq-output2}
	{{{{\bf{P}}_H}\left| {{\psi _r}} \right\rangle } \over {\left\| {{{\bf{P}}_H}\left| {{\psi _r}} \right\rangle } \right\|}} = {{\sum\limits_j {{\lambda _j}{\alpha _j}\left| {{E_j}} \right\rangle } } \over {\left\| {\sum\limits_j {{\lambda _j}{\alpha _j}\left| {{E_j}} \right\rangle } } \right\|}}  .
\end{align}
The task of this quantum judge is to judge whether all eigen values of the given Hamiltonian ${\bf{H}}$ is larger than the threshold $\theta$. More specifically, the quantum judge should output the differentiable states for two different cases. The first case is that all eigen values of the given Hamiltonian ${\bf{H}}$ is larger than $\theta   $ and the second case is that there are states whose corresponding eigen values are smaller than $\theta - \varepsilon$. For convenience of presentation, we define an index $h$ that satisfies
\begin{equation}\label{eq-E-theta}
	{E _j}
	\begin{cases}
		\le \theta - \varepsilon & j \le h ,\\
		\ge \theta    & j > h.
	\end{cases}
\end{equation}
We also assume that the initialization circuit ${{\bf{U}}_I}$ is designed to satisfy that in the second case $\sum\limits_{j \le h} {{{\left| {{\lambda _j}} \right|}^2}}  \ge O\left( {{1 \over { \chi  }}} \right)$.  This is a trivial task since a random circuit can meet this requirement with high probability.

To meet the requirement of output differentiable states in these two cases, we restrain that $\left\{ {{\alpha _j}} \right\}$ in Eq. (\ref{eq-Heaviside-p}) satisfies
\begin{equation}\label{eq-Heaviside-restrain}
	\left| {{\alpha _j}} \right|
	\begin{cases}
		\ge {1 \over 2} & j \le h , \\
		\le O\left( {{1 \over \chi }} \right) & j > h ,
	\end{cases}
\end{equation}
which is a sufficient condition for this requirement when designing the practical quantum Dirac circuit of Eq. (\ref{eq-Heaviside-p}). Now we prove that with the restraint of Eq. (\ref{eq-Heaviside-restrain}), the quantum judge output $\left| 0 \right\rangle $ on the mark qubit with probability no more than $O\left( {{1 \over \chi }} \right)$ in the first case and the output $\left| 0 \right\rangle $ with probability of $O\left( 1 \right)$ in the second case. Here the quantum judge is composed by $O\left( {{1 \over {\sqrt \chi  }}} \right)$ initialization circuits ${{\bf{U}}_I}$ and projector circuits ${{\bf{P}}_H}$.

In the first case, the probability of measuring $\left| 0 \right\rangle $ on the mark qubit from the state ${{\bf{U}}_H}\left| {{\psi _r}} \right\rangle {\left| 0 \right\rangle ^{ \otimes K}}\left| 0 \right\rangle $ is $\sum\limits_j {{{\left| {{\lambda _j}{\alpha _j}} \right|}^2}} $ according to the definition of ${{\bf{U}}_H}$ and $\left| {{\psi _r}} \right\rangle $ from Eq. (\ref{eq-Heaviside-p}) and Eq. (\ref{eq-initial-circuit}) respectively. Based on Eq. (\ref{eq-Heaviside-restrain}), this probability satisfies
\begin{align}
	\label{eq-output-small}
	\sum\limits_j {{{\left| {{\lambda _j}{\alpha _j}} \right|}^2}}  \le O\left( {{1 \over {{\chi ^2}}}} \right)\sum\limits_j {{{\left| {{\lambda _j}} \right|}^2}}  = O\left( {{1 \over {{\chi ^2}}}} \right).
\end{align}
Since we use $O\left( {{1 \over {\sqrt \chi  }}} \right)$ repetitions of ${{\bf{U}}_I}$ and ${{\bf{P}}_H}$, the probability of measuring $\left| 0 \right\rangle $ on the mark qubit in the output state is no more than
\begin{align}
	\label{eq-output3}
	O\left( \chi  \right)\sum\limits_j {{{\left| {{\lambda _j}{\alpha _j}} \right|}^2}}  \le O\left( {{1 \over \chi }} \right),
\end{align}
which is exponentially small. In the second case, we have
\begin{align}
	\label{eq-output-big}
	\sum\limits_j {{{\left| {{\lambda _j}{\alpha _j}} \right|}^2}}  \ge {1 \over 4}\sum\limits_{j \le h} {{{\left| {{\lambda _j}} \right|}^2}}  = O\left( {{1 \over \chi }} \right).
\end{align}
Thus, the amplitude amplification algorithm can amplify the amplitude of $\left| 0 \right\rangle $ on the mark qubit to $O\left( 1 \right)$.

\subsection{Preliminaries of quantum phase estimation}

\begin{figure}[htb]
	\centering
	\includegraphics[width=1\linewidth]{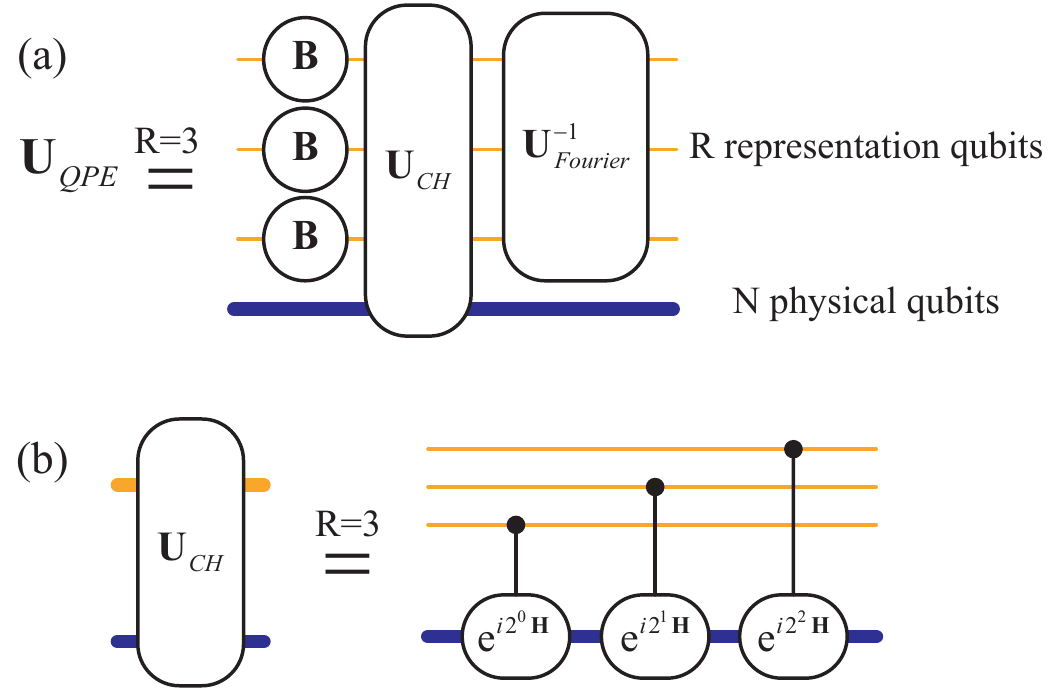}
	\caption{\label{fig-qpe-1}Sketch of quantum phase estimation circuit. (a) The quantum phase estimation circuit of Eq. (\ref{eq-QPE-c}). (b) The circuit of controlled Hamiltonian evolution which achieves Eq. (\ref{ch-result}).}
\end{figure}

The QPE of an eigen state $\left| {{E_j}} \right\rangle $ is exactly given as \cite{Cleve_1998}
\begin{align}
	\label{eq-QPE-e}
	{{\bf{U}}_{QPE}}\left| {{E_j}} \right\rangle {\left| 0 \right\rangle ^{ \otimes R}}{\rm{ = }}{1 \over {{2^R}}}\sum\limits_{x = 0}^{{2^R} - 1} {\sum\limits_{k = 0}^{{2^R} - 1} {{e^{ik{E_j} - 2\pi ik{x \over {{2^R}}}}}\left| {{E_j}} \right\rangle \left| x \right\rangle } }  ,
\end{align}
where ${{\bf{U}}_{QPE}}$ is the QPE circuit and $x$ is a binary number. The QPE circuit is achieved by three parts
\begin{align}
	\label{eq-QPE-c}
	{{\bf{U}}_{QPE}} = \left( {{{\bf{I}}^{ \otimes N}} \otimes {\bf{U}}_{Fourier}^{ - 1}} \right){{\bf{U}}_{CH}}\left( {{{\bf{I}}^{ \otimes N}} \otimes {{\bf{B}}^{ \otimes R}}} \right),
\end{align}
which is shown in Fig. \ref{fig-qpe-1}(a).

Here ${\bf{B}}$ is the quantum Hadamard gate, ${{\bf{U}}_{CH}}$ is a controlled Hamiltonian evolution operator (Fig. \ref{fig-qpe-1}(b)) and ${\bf{U}}_{Fourier}^{ - 1}$ is the inverse quantum Fourier transform. The three circuits can achieve, respectively,
\begin{subequations}\label{eq-three-parts}
	\begin{align}
		{{\bf{B}}^{ \otimes R}}{\left| 0 \right\rangle ^{ \otimes R}} = {\left( {{{\left| 0 \right\rangle  + \left| 1 \right\rangle } \over {\sqrt 2 }}} \right)^{ \otimes R}} = \sqrt {{1 \over {{2^R}}}} \sum\limits_{k = 0}^{{2^R} - 1} {\left| k \right\rangle } ,
	\end{align}
	\begin{align}
		\label{ch-result}
		{{\bf{U}}_{CH}}\sqrt {{1 \over {{2^R}}}} \sum\limits_{k = 0}^{{2^R} - 1} {\left| {{E_j}} \right\rangle \left| k \right\rangle }  = \sqrt {{1 \over {{2^R}}}} \sum\limits_{x = 0}^{{2^R} - 1} {{e^{ik{E_j}}}\left| {{E_j}} \right\rangle \left| k \right\rangle } ,
	\end{align}
	\begin{align}
		{\bf{U}}_{Fourier}^{ - 1}\left| k \right\rangle  = \sqrt {{1 \over {{2^R}}}} \sum\limits_{x = 0}^{{2^R} - 1} {{e^{ - 2\pi ik{x \over {{2^R}}}}}\left| x \right\rangle } .
	\end{align}
\end{subequations}

For convenience, we record the amplitude of ${\left| {{E_j}} \right\rangle \left| x \right\rangle }$ in Eq. (\ref{eq-QPE-e}) as $\kappa \left( {{E_j},x} \right)$, which is
\begin{align}
	\label{eq-kappa}
	\kappa \left( {{E_j},x} \right){\rm{ = }}{1 \over {{2^R}}}\sum\limits_{k = 0}^{{2^R} - 1} {{e^{ik{E_j} - 2\pi ik{x \over {{2^R}}}}}}  .
\end{align}
This is a sum of geometric progression, so when $x$ is not exactly equal to ${2^{R - 1}}{{{E_j}} \over \pi }$, $\kappa \left( {{E_j},x} \right)$ can be written as
\begin{align}
	\label{eq-kappa-simple}
	\kappa \left( {{E_j},x} \right){\rm{ = }}{1 \over {{2^R}}}{{1 - {e^{i{2^R}{E_j} - 2\pi ix}}} \over {1 - {e^{i{E_j} - 2\pi i{x \over {{2^R}}}}}}} .
\end{align}
When $\left| {{{{E_j}} \over {2\pi }} - {x \over {{2^R}}}} \right| \le {1 \over {{2^{R + 1}}}}$, it can be proved that the modulus of $\kappa \left( {{E_j},x} \right)$ satisfies
\begin{align}
	\label{eq-kappa-proof1}
	\left| {\kappa \left( {{E_j},x} \right)} \right| & = \left| {{{\sin \left( {{2^{R - 1}}{E_j} - \pi x} \right)} \over {{2^R}\sin \left( {{{{E_j}} \over 2} - \pi {x \over {{2^R}}}} \right)}}} \right| \notag \\ & \ge {1 \over {{2^R}}}{{\left| {\sin \left( {{2^{R - 1}}{E_j} - \pi x} \right)} \right|} \over {\pi \left| {{{{E_j}} \over {2\pi }} - {x \over {{2^R}}}} \right|}} \notag \\ & \ge {1 \over {{2^R}}}{{{2^{R + 1}}\left| {{{{E_j}} \over {2\pi }} - {x \over {{2^R}}}} \right|} \over {\pi \left| {{{{E_j}} \over {2\pi }} - {x \over {{2^R}}}} \right|}} \notag \\ & = {2 \over \pi }.
\end{align}
We can define a classical function ${\rm{n}}\left( y \right)$ to output the nearest binary number on $R$ digits of ${{2^R}y}$, where $y$ is in the range of $\left[ {0,1 - {1 \over {{2^R}}}} \right]$. It can be easily seen that
\begin{align}
	\label{eq-n-diff}
	\left| {{{n\left( y \right)} \over {{2^R}}} - y} \right| \le {1 \over {{2^{R + 1}}}}.
\end{align}
Without losing the generality, we assume ${{E_j}}$ is in the range of $\left[ {0,2\pi  - {{2\pi } \over {{2^R}}}} \right]$, and then Eq. (\ref{eq-kappa-proof1}) becomes
\begin{align}
	\label{eq-kappa-proof2}
	\left| {\kappa \left( {{E_j},n\left( {{{{E_j}} \over {2\pi }}} \right)} \right)} \right| \ge {2 \over \pi }.
\end{align}
Eq. (\ref{eq-kappa-proof2}) shows that the QPE method can output a binary representation of ${n\left( {{{{E_j}} \over {2\pi }}} \right)}$ with the amplitude larger than ${2 \over \pi }$. Thus, we can obtain the value of ${{{{E_j}} \over {2\pi }}}$ with an error lower than $O\left( {{1 \over {{2^R}}}} \right)$.

For the convenience of our later proofs on quantum Heaviside circuit, here we give a lower bound of the modulus of $\kappa \left( {{E_j},x} \right)$. When $\left| {{{{E_j}} \over {2\pi }} - {x \over {{2^R}}}} \right| \le {1 \over {S{2^{R + 1}}}}$ and $S \ge 2$ \cite{Poulin_2009}, the lower bound of the modulus of $\kappa \left( {{E_j},x} \right)$ is
\begin{align}
	\label{eq-kappa-low}
	\left| {\kappa \left( {{E_j},x} \right)} \right| & = {1 \over {{2^R}}}\left| {\sum\limits_{k = 0}^{{2^R} - 1} {{e^{ik{E_j} - 2\pi ik{x \over {{2^R}}}}}} } \right| \notag \\ & \ge {1 \over {{2^R}}}\left| {\sum\limits_{k = 0}^{{2^R} - 1} {\cos \left[ {2\pi k\left( {{{{E_j}} \over {2\pi }} - {x \over {{2^R}}}} \right)} \right]} } \right| \notag \\ & \ge {1 \over {{2^R}}}\sum\limits_{k = 0}^{{2^R} - 1} {\cos \left( {{\pi \over S}} \right)}  \notag \\ & = \cos \left( {{\pi \over S}} \right) \notag \\ & \ge 1 - {{\pi}^2 \over {2{S^2}}}.
\end{align}

\subsection{The first strategy in the quantum Heaviside circuit}

\begin{figure}[htb]
	\centering
	\includegraphics[width=1\linewidth]{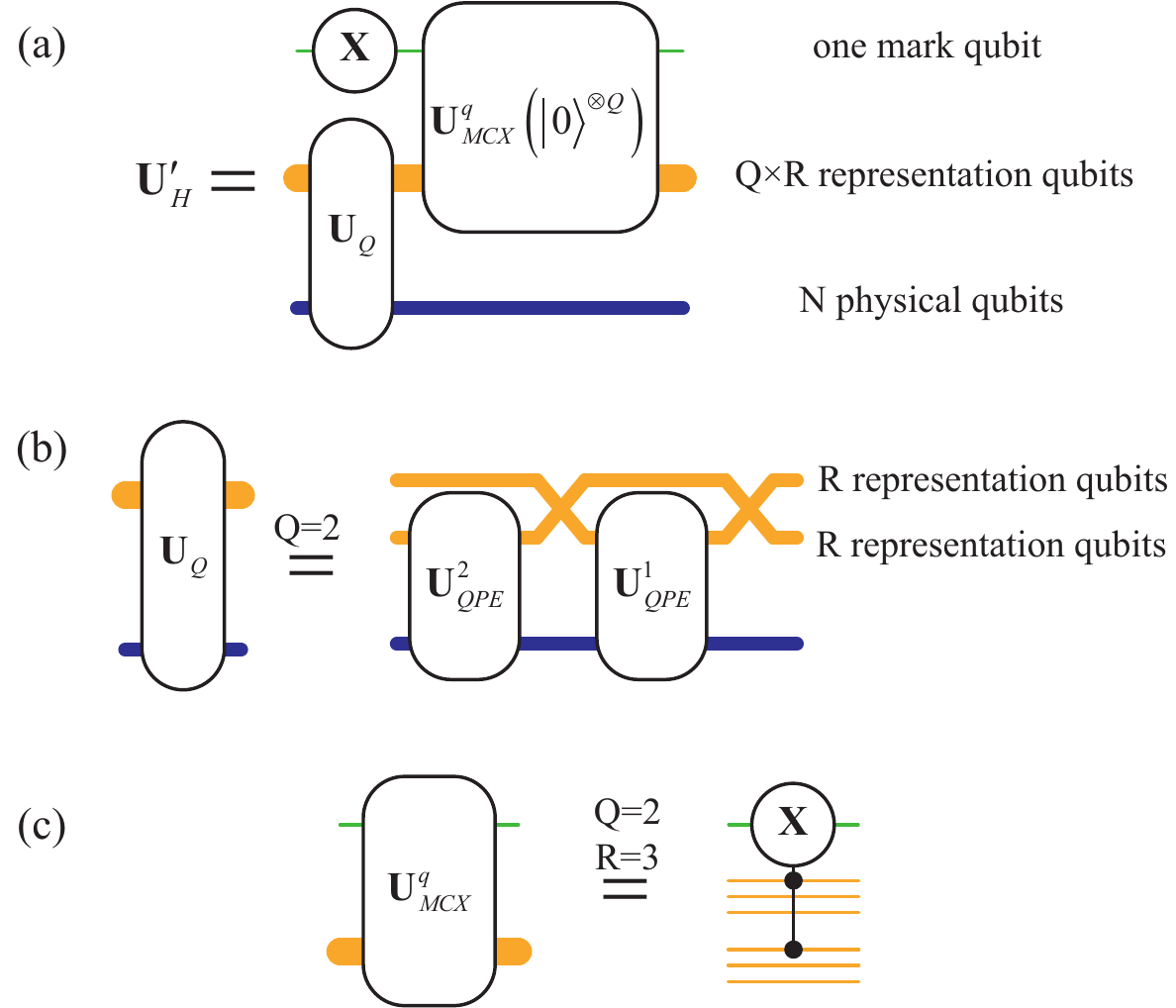}
	\caption{\label{fig-heaviside-1}(a) guide to construct the primary quantum Heaviside circuit of Eq. (\ref{eq-Heaviside-1}). (a) The primary design of quantum Heaviside circuit of Eq. (\ref{eq-Heaviside-1}). This circuit can successfully filter out the states in the bad subspace, but it requires huge amount of qubits and cannot preserve the good states. (b) The circuit to apply the quantum phase estimation to $Q$ parts of representation qubits as given in Eq. (\ref{eq-q-QPE}). (c) The circuit to apply NOT gate to the mark qubit controlled by all first qubits of all $Q$ parts of representation qubits.}
\end{figure}

As we have shown in Eq. (\ref{eq-QPE-e}), the implementation of QPE circuit requires $N$ physical qubits and $R$ representation qubits. The first strategy in QHC is to implement $Q$ QPE circuits on the same physical qubits and $Q \times R$ different representation qubits. This circuit is defined as
\begin{align}
	\label{eq-q-QPE}
	{\bf{U}}_{Q} = \prod\limits_q^Q {{\bf{U}}_{QPE}^q},
\end{align}
where ${{\bf{U}}_{QPE}^q}$ is the extension of QPE circuit on the $q$-th part of representation qubits, as shown in Fig. \ref{fig-heaviside-1}(b). Without losing the generality, we assume that all eigen values are in the range of $\left[ {0,2\pi  - {{2\pi } \over {{2^R}}}} \right]$ and the given threshold $\theta $ equals $\pi $.

Ignoring the huge amount of qubits it needs, we now give the primary design of quantum Heaviside circuit
\begin{align}
	\label{eq-Heaviside-1}
	{{\bf{U'}}_H} = {\bf{U}}_{MCX}^q\left( {{{\left| 0 \right\rangle }^{ \otimes Q}}} \right)\left( {{{\bf{U}}_Q} \otimes {\bf{I}}} \right)\left( {{{\bf{I}}^{ \otimes N + QR}} \otimes {\bf{X}}} \right),
\end{align}
where ${\bf{U}}_{MCX}^q$ is a NOT gate on the mark qubit controlled by $Q$ first (the highest) qubits of $Q$ parts of representation qubits. This primary design of quantum Heaviside circuit is shown in Fig. \ref{fig-heaviside-1}. The effect of this circuit can be shown as
\begin{align}
	\label{eq-Heaviside-effect}
	{{\bf{U'}}_H}\left| {{E_j}} \right\rangle {\left| 0 \right\rangle ^{ \otimes {QR+1}}} & {\rm{ = }}\left| {{E_j}} \right\rangle {\left( {\sum\limits_{x = 0}^{{2^{R-1}} - 1} {\kappa \left( {{E_j},x} \right)\left| x \right\rangle } } \right)^{ \otimes Q}}\left| 0 \right\rangle \notag \\ & + \left| {{E_j}} \right\rangle \left| {rest} \right\rangle \left| 1 \right\rangle  .
\end{align}

This quantum Heaviside circuit can successfully filter out the states in the bad subspace while it cannot ensure to preserve the good states. This is equivalent to satisfying the second requirement of Eq. (\ref{eq-Heaviside-restrain}) and ignoring the first requirement. It can be proved below when
\begin{subequations}\label{eq-reatrain-QR}
	\begin{align}\label{eq-reatrain-R}
		\varepsilon  \ge {2 \pi \over {{2^{R - 1}}}},
	\end{align}
	\begin{align}\label{eq-reatrain-Q}
		Q = O\left( N \right) .
	\end{align}
\end{subequations}
From Eq. (\ref{eq-reatrain-R}) and Eq. (\ref{eq-E-theta}), we can obtain that $n\left( {{{{E_j}} \over {2\pi }}} \right) > 2^{R-1}$ for $j > h$. This means that the qubit with the highest order in the binary representation of $n\left( {{{{E_j}} \over {2\pi }}} \right)$ is $\left| 1 \right\rangle $. Based on Eq. (\ref{eq-kappa-proof2}) and Eq. (\ref{eq-Heaviside-effect}), the amplitude of the mark qubit being $\left| 0 \right\rangle $ satisfies
\begin{align}
	\label{eq-4132-1}
	{\left| {{\alpha _j}} \right|^2} & = {\left( {\sum\limits_{x = 0}^{{2^{R - 1}} - 1} {{{\left| {\kappa \left( {{E_j},x} \right)} \right|}^2}} } \right)^Q} \notag \\ & \le {\left( {1 - {{\left| {\kappa \left( {{E_j},n\left( {{{{E_j}} \over {2\pi }}} \right)} \right)} \right|}^2}} \right)^Q} \notag \\ & \le {\left( {1 - {4 \over {{\pi ^2}}}} \right)^Q}
\end{align}
Since $N = O\left( {\log \chi } \right)$, we can easily find that $Q$ satisfies Eq. (\ref{eq-reatrain-Q}) and meets the second requirement of Eq. (\ref{eq-Heaviside-restrain}), which is $\left| {{\alpha _j}} \right| \le {\rm{O}}\left( {{1 \over \chi }} \right)$.

\subsection{The second strategy in the quantum Heaviside circuit}

Quantum Heaviside circuit using only the first strategy may unfortunately filter out the good states, and the second strategy is to implement a batch of QPE circuits ${{\bf{U}}_{Q}}$ for $W$ different Hamiltonian matrices, which are expressed as $\left\{ {{{\bf{H}}_w}} \right\}:{{\bf{H}}_w} = {{\bf{H}}_0} + {w \over W}{\pi \over {{2^{R - 1}} }}$. Now we prove that the good states of ${\bf{H}}_0$ can survive at least once in the $W$ quantum Heaviside circuit. Suppose that the eigen value of a good state of ${\bf{H}}_0$ is ${E}_q$, then there is a set of ${E}_q$ corresponding to the set of ${{{\bf{H}}_w}}$ recorded as
\begin{align}
	\label{eq-E-set}
	\left\{ {{E_{q,w}}} \right\}:{E_{q,w}} = {E_q} + {w \over W}{\pi \over {{2^{R - 1}} }}.
\end{align}
Note that ${E}_q$ is the eigen value of a good state which means that all ${E_{s,w}}$ satisfy
\begin{align}
	\label{eq-E-big}
	n\left( {{{{E_{q,w}}} \over {2\pi }}} \right) < {2^{R - 1}}.
\end{align}

Since the interval of ${E_{s,w}}$ is ${1 \over {W{2^{R - 1}}\pi }}$, there will be at least one ${w_0}$ that satisfies
\begin{align}
	\label{eq-w-good}
	\left| {{{{E_{q,{w_0}}}} \over {2\pi }} - {1 \over {{2^R}}}n\left( {{{{E_{q,{w_0}}}} \over {2\pi }}} \right)} \right| \le {1 \over {W{2^{R+1}} }}.
\end{align}
Substituting this equation into Eq. (\ref{eq-kappa-low}), we can get the amplitude immediately
\begin{align}
	\label{eq-w-good-a}
	\left| {\kappa \left( {{E_{q,{w_0}}},n\left( {{{{E_{q,{w_0}}}} \over {2\pi }}} \right)} \right)} \right| \ge 1 - {{\pi}^2 \over {2{W^2}}}.
\end{align}
Note that the amplitude of the mark qubit being $\left| 0 \right\rangle $ after the quantum Heaviside circuit satisfies
\begin{align}
	\label{eq-4132-2}
	{\left| {{\alpha _q}} \right|^2}{\rm{ = }}{\left[ {\sum\limits_{x = 0}^{{2^R-1} - 1} {{{\left| {\kappa \left( {{E_q},x} \right)} \right|}^2}} } \right]^Q} \ge {\left[ {{{\left| {\kappa \left( {{E_q},n\left( {{{{E_q}} \over {2\pi }}} \right)} \right)} \right|}^2}} \right]^Q}.
\end{align}
To meet the first requirement in Eq. (\ref{eq-Heaviside-restrain}), we obtain
\begin{align}
	\label{eq-W}
	{W = O\left( {\sqrt Q } \right) = O\left( {\sqrt N } \right)}.
\end{align}
Since the condition for the establishment of Eq. (\ref{eq-4132-1}) still holds, which is ${n\left( {{{{E_{s,w}}} \over {2\pi }}} \right) > {2^{R - 1}}}$, all $W$ quantum Heaviside circuits can filter out the bad states, whose eigen values are larger than ${\pi }$.

\begin{figure}[htb]
	\centering
	\includegraphics[width=1\linewidth]{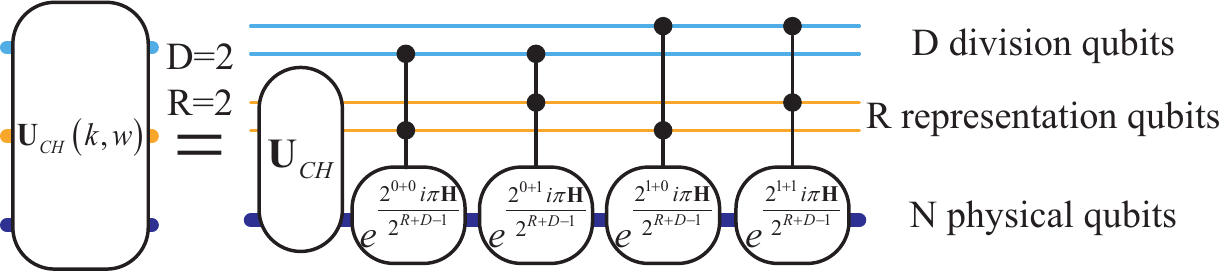}
	\caption{\label{fig-chd}The circuit to achieve the result of Eq. (\ref{eq-Kh-d}). This circuit contains $D \times R$ controlled Hamiltonian evolution operator in addition to the ${{\bf{U}}_{CH}}$.}
\end{figure}

The classical part of this method can be replaced by a quantum search algorithm of ${{w_0}}$ that satisfies Eq. (\ref{eq-w-good}). It can speed up this method by ${O\left( {\sqrt W } \right) = O\left( {{N^{{1 \over 4}}}} \right)}$. Now we define a shifted QPE circuit with $D$ extra division qubits that can achieve
\begin{align}
	\label{eq-sQPE-r}
	& {{\bf{U}}_{sQPE}}\left| {{E_j}} \right\rangle {\left| 0 \right\rangle ^{ \otimes R}}{\left| 0 \right\rangle ^{ \otimes D}}{\rm{ = }} \notag \\ & {1 \over {{2^R}}}\sum\limits_{x = 0}^{{2^R} - 1} {\sum\limits_{k = 0}^{{2^R} - 1} {\sum\limits_{w = 0}^{{2^D} - 1} {{e^{ik{E_j} + 2\pi i{kw \over {{2^{R + D}}}} - 2\pi ik{x \over {{2^R}}}}}\left| {{E_j}} \right\rangle \left| x \right\rangle } } } \left| w \right\rangle .
\end{align}
The operator ${{\bf{U}}_{sQPE}}$ can be easily constructed by combining the QPE circuit shown in Eq. (\ref{eq-QPE-c}) with a controlled Hamiltonian evolution operator. To distinguish it from the controlled Hamiltonian evolution operator in Eq. (\ref{eq-QPE-c}), we record the two operators as $ {{\bf{U}}_{CH}}\left( k \right)$ and ${{\bf{U}}_{CH}}\left( {k,w} \right)$ which can achieve respectively
\begin{subequations}\label{eq-Kh}
	\begin{align}
		{{\bf{U}}_{CH}}\left( k \right)\left| {{E_j}} \right\rangle \left| k \right\rangle  = {e^{ik{E_j}}}\left| {{E_j}} \right\rangle \left| k \right\rangle ,
	\end{align}
	\begin{align}
		\label{eq-Kh-d}
		{{\bf{U}}_{CH}}\left( {k,w} \right)\left| {{E_j}} \right\rangle \left| k \right\rangle \left| w \right\rangle  = {e^{ik{E_j} + 2\pi i{{kw} \over {{2^{R + D}}}}}}\left| {{E_j}} \right\rangle \left| k \right\rangle \left| w \right\rangle,
	\end{align}
\end{subequations}
where ${{\bf{U}}_{CH}}\left( k \right)$ is the controlled Hamiltonian evolution used in Eq. (\ref{eq-QPE-c}). The quantum circuit ${{\bf{U}}_{CH}}\left( {k,w} \right)$ is represented in Fig. \ref{fig-chd}.

The construction of shifted QPE is to replace the controlled Hamiltonian evolution operator in Eq. (\ref{eq-QPE-c}) with ${{\bf{U}}_{CH}}\left( {k,w} \right)$, where the division qubits are initialized to $\sum\limits_{w = 0}^{{2^D} - 1} {\left| w \right\rangle } $. The shifted QPE operator is shown as
\begin{align}
	\label{eq-sQPE-c}
	{{\bf{U}}_{sQPE}} = & \left( {{{\bf{I}}^{ \otimes N}} \otimes {\bf{U}}_{Fourier}^{ - 1} \otimes {{\bf{I}}^{ \otimes D}}} \right){{\bf{U}}_{CH}}\left( {k,w} \right) \notag \\ & \left( {{{\bf{I}}^{ \otimes N}} \otimes {{\bf{B}}^{ \otimes R}} \otimes {{\bf{I}}^{ \otimes D}}} \right).
\end{align}
Note that the quantum Fourier transform still works on the $R$ representation qubits.

The simple replacement of ${{\bf{U}}_{CH}}\left( k \right)$ with ${{\bf{U}}_{CH}}\left( {k,w} \right)$ reduces the lower bound on probability of outputting qualified states from $O\left( 1 \right)$ to $O\left( {{1 \over W}} \right)$. This process can be seen as to perform ${{\bf{U'}}_H}$ for $W$ different ${{\bf{H}}_w}$'s in parallel, compared with serial performing ${{\bf{U'}}_H}$ without the division qubits. The parallel quantum Heaviside circuit is recorded as
\begin{align}
	\label{eq-p-Heaviside}
	{{\bf{U'}}_{p - H}} = {\bf{U}}_{MCX}^q\left( {{{\bf{U}}_{sQ}} \otimes {{\bf{I}}}} \right),
\end{align}
where ${\bf{U}}_{MCX}^q$ is a NOT gate on the mark qubit controlled by $Q$ first qubits of $Q$ parts of representation qubits. The ${{{\bf{U}}_{sQ}}}$ is defined as
\begin{align}
	\label{eq-q-sQPE}
	{{{\bf{U}}_{sQ}} = \prod\limits_q^Q {{\bf{U}}_{sQPE}^q} },
\end{align}
which is similar to Eq. (\ref{eq-q-QPE}).

For the consistency of the definition of quantum Heaviside circuit, the probability of outputting qualified states should be increased to  $O\left( 1 \right)$ by the fixed-point search. This version of quantum Heaviside circuit is defined as
\begin{align}
	\label{eq-Heaviside-2}
	{{\bf{U''}}_H} = {\bf{F}}\left[ {{{{\bf{U'}}}_{p - H}},{{\bf{I}}^{ \otimes N + QR + D}} \otimes \left( {\left| 0 \right\rangle \left\langle 0 \right|} \right)} \right],
\end{align}
where ${\left| 0 \right\rangle \left\langle 0 \right|}$ is a projector to ${\left| 0 \right\rangle }$ on the mark qubit. Compared with the previous method with classical search, this method is more elegant in the sense that it is a complete quantum algorithm and has a quadratic speedup in the search of Hamiltonian set $\left\{ {{{\bf{H}}_w}} \right\}$. But in practice, we still adopt the previous design of quantum Heaviside circuit of Eq. (\ref{eq-Heaviside-1}) combining with the classical search of $\left\{ {{{\bf{H}}_w}} \right\}$. The reason is that the speedup scale is only $O\left( {\sqrt W } \right) = O\left( {{N^{{1 \over 4}}}} \right)$ according to Eq. (\ref{eq-W}). Furthermore, this quantum search greatly increases the difficulty of designing and simulating the quantum judge.

\subsection{The third strategy in quantum Heaviside circuit}

We have introduced the freezing operator to reduce the number of required auxiliary qubits in the construction of quantum Dirac circuit. One should note that the initial state and the target state are both $\left| 0 \right\rangle $ on the mark qubit in quantum Dirac circuit, where only one freezing operator is needed to freeze $\left| 1 \right\rangle $ on the mark qubit. In quantum Heaviside circuit of Eq. (\ref{eq-Heaviside-1}), the initial state is ${\left| 0 \right\rangle ^{ \otimes R}}$ on the representation qubits, while the target state is $\left| 0 \right\rangle $ on the first (the highest order) qubit of the representation qubits. Now we need two freezing operators to freeze the states orthogonal to ${\left| 0 \right\rangle ^{ \otimes R}}$ on the representation qubits and the state $\left| 0 \right\rangle $ on the mark state respectively. The first freezing operator is to extract the target state on the mark qubits and the second one guarantees that the QPE circuit always starts on the right state.

\begin{figure}[htb]
	\centering
	\includegraphics[width=0.9\linewidth]{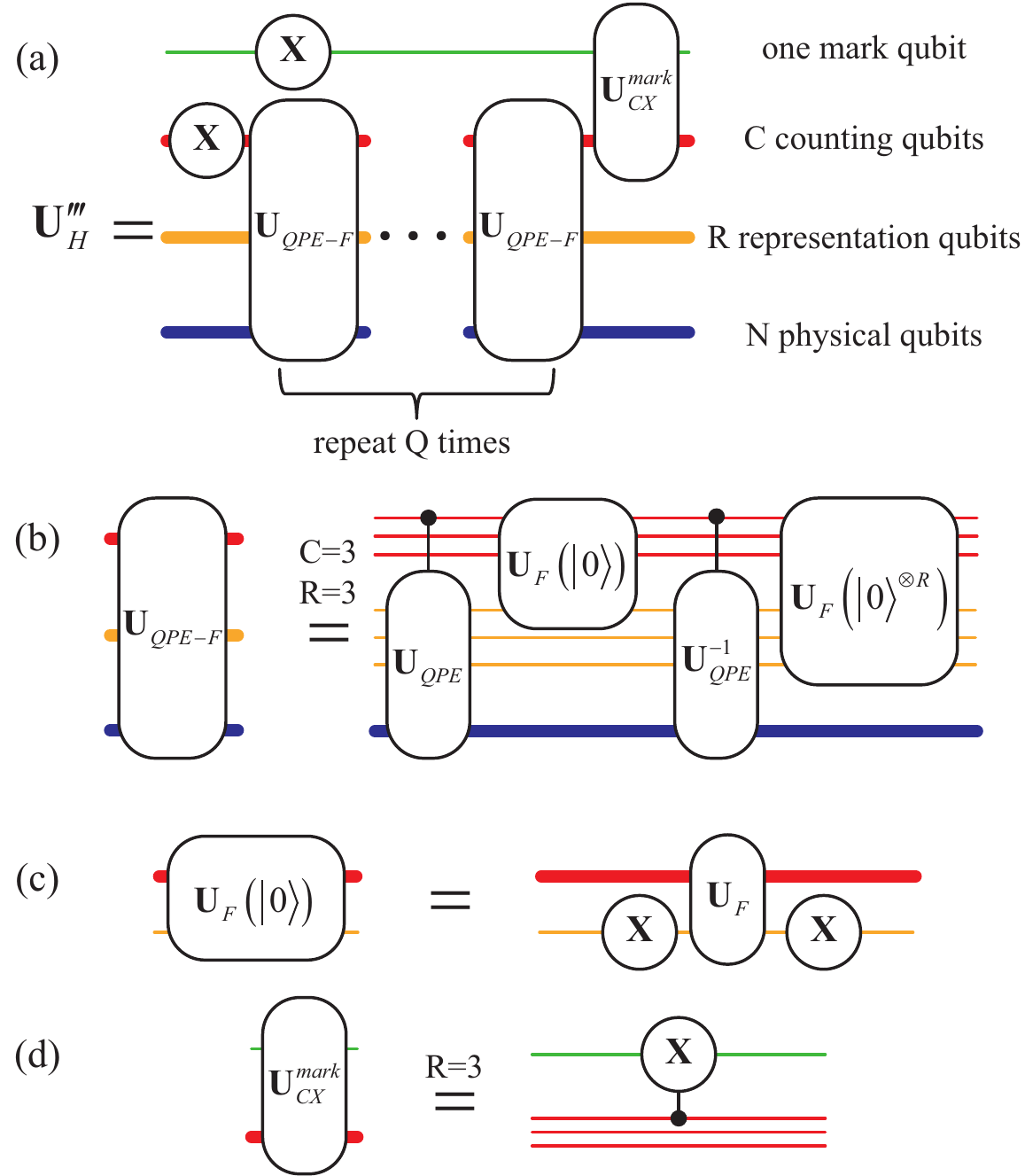}
	\caption{\label{fig-heaviside-3}(a) guide to construct the quantum Heaviside circuit of Eq. (\ref{eq-Heaviside-3}). (a) The design of quantum Heaviside circuit of Eq. (\ref{eq-Heaviside-3}) used for the quantum judge. (b) The repeating units to perform the quantum phase estimation and its inverse $Q$ times. (c) The freezing operator whose target state is $\left| 0 \right\rangle $ on the first qubit of the representation qubits. (d) The NOT gate on the mark qubit controlled by the first counting qubit. }
\end{figure}

The primary design of quantum Heaviside circuit of Eq. (\ref{eq-Heaviside-1}) requires $Q \times R$ representation qubits. Combining with the freezing circuit, we can perform $Q$ times of QPE on $R$ representation qubits and $O\left( {\log Q} \right)$ counting qubits instead of $Q \times R$ representation qubits. This new circuit is
\begin{align}
	\label{eq-Heaviside-3}
	{{\bf{U'''}}_H} = {\bf{U}}_{CX}^{mark}\left[ {{{\left( {{{\bf{U}}_{QPE - F}}} \right)}^Q} \otimes {\bf{I}}} \right]\left( {{{\bf{I}}^{ \otimes N + R}} \otimes {{\bf{X}}^{ \otimes C + 1}}} \right).
\end{align}
${\bf{U}}_{CX}^{mark}$ is a NOT gate on the mark qubit controlled by the first (the highest order) qubit of the $C$ counting qubits. ${{{\bf{U}}_{QPE - F}}}$ is expressed as
\begin{align}
	\label{eq-QPE-F}
	{{\bf{U}}_{QPE - F}} = \left( {{{\bf{I}}^{ \otimes N}} \otimes {{\bf{U}}_{{F_2}}}} \right)C{\bf{U}}_{QPE}^{ - 1}\left( {{{\bf{I}}^{ \otimes N}} \otimes {{\bf{U}}_{{F_1}}}} \right)C{{\bf{U}}_{QPE}},
\end{align}
where $C{{\bf{U}}_{QPE}}$ and $C{\bf{U}}_{QPE}^{ - 1}$ are the QPE circuit and its inverse controlled by the counting qubits, respectively. ${{{\bf{U}}_{{F_1}}}}$ and ${{{\bf{U}}_{{F_2}}}}$ are freezing operators with target states of $\left| 0 \right\rangle $ on the first qubit of the representation qubits and ${\left| 0 \right\rangle ^{ \otimes R}}$ on the representation qubits, respectively. The sketch of this construction is depicted in Fig. \ref{fig-heaviside-3}.

\subsection{Construction and gate complexity of quantum judge}

The quantum judge can be constructed by the quantum Heaviside circuit as
\begin{align}
	\label{eq-judge}
	{\bf{J}} = \left[ {{{\bf{I}}^{ \otimes N + R + C}} \otimes \left( {\left| 0 \right\rangle \langle 0|} \right)} \right]{\bf{F}}\left[ {{{{\bf{U'''}}}_H},{{\bf{I}}^{ \otimes N + R + C}} \otimes \left( {\left| 0 \right\rangle \langle 0|} \right)} \right].
\end{align}
One must note that the quantum judge is not a unitary operator as it contains a projector that is achieved by measurements. The process using this quantum judge to solve the lowest eigen value of the given Hamiltonian matrix $\bf{H}$ contains three steps. The first step is to normalize the Hamiltonian matrix to make sure that all eigen values are in the range of $\left[ {0,{{{2^M} - 1} \over {{2^{M - 1}}}}\pi } \right]$. Then the second step is to use the quantum judge to decide whether all Hamiltonian in the set of $\left\{ {{{\bf{H}}_w}} \right\}$ do not output ${\left| 0 \right\rangle }$ on the mark qubit with a probability of $O\left( 1 \right)$, which indicates that no eigen value of the given Hamiltonian is lower than the given threshold $\theta $. The third step is to use dichotomy to obtain the lowest eigen value with an error lower than $\varepsilon $ in $O\left( {\log {1 \over \varepsilon }} \right)$ iterations.

The complexity analysis can be easily done from the sketch of quantum Heaviside circuit in Fig. \ref{fig-heaviside-3}. The number of auxiliary qubits is $R+C+1$, which is rewritten as $O\left( {\log N + \log {1 \over \varepsilon }} \right)$ based on $C = O\left( {\log Q} \right)$. The analysis of gate complexity is similar to the case of quantum Dirac circuit. The only new circuit is the QPE circuit, whose gate complexity is $\tilde O\left( {{\Lambda  \over \varepsilon }} \right)$ \cite{Cleve_1998}. Here we stress that the (inverse) Hamiltonian evolution ${\left( {{e^{iH}}} \right)^t}$ is achieved by repeating ${{e^{iH}}}$ $t$ times instead of simulating the Hamiltonian for time $t$, which can avoid the error accumulation from the imperfect Hamiltonian simulation \cite{Poulin_2009}. As no other new circuit is introduced in quantum Heaviside circuit of Eq. (\ref{eq-Heaviside-3}), the gate complexity of the quantum judge is
\begin{align}
	\label{eq-gate-h-3}
	\tilde O\left( {{\Lambda  \over {\varepsilon \sqrt \chi  }} + {\Phi  \over {\sqrt \chi  }}} \right).
\end{align}